\documentclass[12pt,preprint]{aastex}

\slugcomment{}

\shorttitle{Globular Clusters Relative Ages}
\shortauthors{Mar\' \i n-Franch et al.}

\begin{document}

\title{The ACS Survey of Galactic Globular Clusters. VII.\thanks{
Based on observations with the NASA/ESA {\it Hubble Space Telescope},
obtained at the Space Telescope Science Institute, which is operated
by AURA, Inc., under NASA contract NAS 5-26555, under program
GO-10775 (PI: Sarajedini).
}
\\ Relative Ages}

\author{Antonio Mar\' \i n-Franch}
\affil{Instituto de Astrof\'\i sica de Canarias, V\'\i a L\'actea s/n, E-38200 La Laguna, Spain, and 
Department of Astronomy, University of Florida, 211 Bryant Space Science Center, Gainesville, FL 32611}
\email{amarin@iac.es}

%%%% Relative Ages core group

\author{Antonio Aparicio}
\affil{University of La Laguna and Instituto de Astrof\'\i sica de Canarias, E-38200 La Laguna, Spain
\email{antapaj@iac.es} }

\author{Giampaolo Piotto}
\affil{Dipartimento di Astronomia, Universit\`{a} di Padova, 35122 Padova, Italy 
\email{giampaolo.piotto@unipd.it} }

\author{Alfred Rosenberg}
\affil{Instituto de Astrof\'\i sica de Canarias, V\'\i a L\'{a}ctea s/n, E-38200 La Laguna, Spain
\email{alf@iac.es} }

\author{Brian Chaboyer}
\affil{Department of Physics and Astronomy, Dartmouth College, 6127 Wilder Laboratory, Hanover, NH 03755 
\email{chaboyer@heather.dartmouth.edu} }

%%%% Rest of GCTreasury members

\author{Ata Sarajedini}
\affil{Department of Astronomy, University of Florida, 211 Bryant Space Science Center, Gainesville, FL 32611
\email{ata@astro.ufl.edu} }

\author{Michael Siegel}
\affil{University of Texas, McDonald Observatory, 1 University Station, C1402, Austin TX, 78712
\email{siegel@astro.as.utexas.edu} }

\author{Jay Anderson}
\affil{Space Telescope Science Institute, 3700 San Martin Drive, Baltimore MD 21218
\email{jaynder@stsci.edu} }

\author{Luigi R. Bedin}
\affil{Space Telescope Science Institute, 3700 San Martin Drive, Baltimore MD 21218
\email{bedin@stsci.edu} }

\author{Aaron Dotter}
\affil{Department of Physics and Astronomy, Dartmouth College, 6127 Wilder Laboratory, Hanover, NH 03755 
\email{Aaron.L.Dotter@dartmouth.edu} }

\author{Maren Hempel}
\affil{Department of Astronomy, University of Florida, 211 Bryant Space Science Center, Gainesville, FL 32611
\email{hempel@astro.ufl.edu} }

\author{Ivan King}
\affil{Dept. of Astronomy, Univ. of Washington, Box 351580, Seattle, WA 98195-1580
\email{king@astro.washington.edu} }

\author{Steven Majewski}
\affil{Dept. of Astronomy, University of Virginia, P.O. Box 400325, Charlottesville, VA 22904-4325
\email{srm4n@virginia.edu} }

\author{Antonino P. Milone}
\affil{Dipartimento di Astronomia, Universit\`{a} di Padova, 35122 Padova, Italy 
\email{antonino.milone@oapd.inaf.it}}

\author{Nathaniel Paust}
\affil{Space Telescope Science Institute, 3700 San Martin Drive, Baltimore MD 21218
\email{paust@stsci.edu} }

\author{I. Neill Reid}
\affil{Space Telescope Science Institute, 3700 San Martin Drive, Baltimore MD 21218
\email{inr@stsci.edu} }

\begin{abstract}

The ACS Survey of Galactic Globular Clusters is a
$Hubble$ $Space$ $Telescope$ $(HST)$ Treasury program
designed to provide a new large, deep and homogeneous
photometric database. Based on observations from this program, we have
measured precise relative ages for a sample of 64 Galactic globular
clusters by comparing the relative position of the clusters' main 
sequence turn offs, using main$-$sequence fitting
to cross$-$compare clusters within the sample.
This method provides relative ages to a formal precision 
of 2-7\%. We demonstrate that the calculated relative ages 
are independent of the choice of theoretical model.
We find that the Galactic globular cluster sample can be divided into two
groups -- a population of old clusters with an age dispersion of
$\sim$5\% and no age$-$metallicity relation, and a group of
younger clusters with an age$-$metallicity relation
similar to that of the globular clusters associated with the
Sagittarius dwarf galaxy. 

These results are consistent with the Milky Way halo having formed in
two phases or processes. The first one would be compatible with a rapid ($<$0.8 Gyr) assembling process of the halo,
in which the clusters in the old group were formed. Also these clusters could have been formed before reionization in dwarf galaxies that would later merge to build the Milky Way halo as predicted by $\Lambda$CDM cosmology. However, the galactocentric metallicity gradient shown by these clusters seems difficult to reconcile with the latter. As for the younger clusters, it is very tempting to argue that their origin is related to their formation within Milky Way satellite galaxies that were later accreted, but the origin of the age$-$metallicity relation remains unclear. 

\end{abstract}

\keywords{Galaxy: evolution --- Galaxy: formation --- globular clusters: general}

\section{Introduction}

One of the keys to understanding the structure and evolution
of the Milky Way is the ability to divide its stars and clusters into
separate Galactic Populations as first hinted at by
\citet{O26} and formally introduced by \citet{B44}. In the early 1950s, Sandage published 
the first deep color$-$magnitude diagrams (CMDs) of
the northern Galactic globular clusters (GGCs) M3 and M92 \citep{S53, A53} and
discovered the faint sequence of stars similar to the main sequence of Population I.
It was almost immediately realized that GGCs are ancient compared to the stars
in the Solar Neighborhood.

Over the last half century, it has become clear that GGCs are astronomical
fossils providing important information on the formation and evolution
of the Milky Way. One characteristic of GGCs that makes them so valuable
is that they are the oldest Galactic objects for which
reliable ages can be measured.

GGC ages are of great interest to a variety of cosmological and comogonical issues
since their {\it absolute} ages place a lower limit on the age of the 
Universe. Unfortunately, absolute age measurements are still compromised by
a number of uncertainties, particularly GGC distances and 
foreground reddenings as well as metallicities, bolometric corrections and some poorly known aspects of the stellar evolution. Relative ages are almost (although not totally) free of these effects and can still reveal fundamental information about Galaxy formation mechanisms and time scales. For this reason they have become of paramount importance in the context of the study of GGC.

The study of GGC ages has evolved apace with improvements in instrumentation, in our
theoretical understanding of stellar evolution and with the advance of computational facilities.
Reliable age determination requires deep photometry that reaches at least the main$-$sequence turn off
(MSTO),
together with a good understanding of the stellar evolution theoretical modeling.
During the first
half of the 20$^{\rm th}$ century, the available instrumentation did not
allow observations deeper than the horizontal branch of most GGCs.
Over the past three decades, however, low quantum efficiency
non$-$linear photographic imaging has given way to digital imaging using
high quantum efficiency linear CCDs.  This has resulted in ever
deeper and more precise cluster CMDs, 
that reached the MSTO and allowed the application of age dating techniques. 
First relative age studies were carried out by \citet{G85}, \citet{P87}, \citet{SK89} and
\citet{SD90}, as representative examples.
However, uncertainties in both the input physics \citep{Ch98} and  
the color$-$T$_{\rm eff}$ transformation of the theoretical models
into the observational plane \citep{B98} constitute a 
significant source of errors in the final relative age determination, particularly when 
ages of clusters with different metallicities are compared.
In the last decade, many efforts have 
been made in GGC relative ages, improving our understanding with each technological or
theoretical advance \citep[e. g. ][]{Ch96, R96, Stet96, SW98, B98, R99, Van00, DeA05}.

Relative ages can be estimated to high precision
by measuring the position of the MSTO relative 
to CMD features that have little or no age dependence \citep{Stet96, SarCha97}.
However, a key aspect of any relative age experiment is having a large homogenous database
of globular cluster photometry.  Such databases have only recently
appeared in the literature but have significantly advanced our
understanding of Milky Way formation and evolution. The largest
dedicated efforts are those of \citet{R00a, R00b}, a
ground-based survey of
35 nearby GGCs \citep{R99}, and \citet{DeA05}, an analysis of 55 clusters
from \citet{P02}'s HST snapshot catalog.

These and other recent relative age studies \citep{B98, SW98, R99, Van00, DeA05} consistently
point towards a scenario in which metal$-$poor clusters 
are largely coeval but
metal$-$rich clusters show some age dispersion, $\sim$1.0 Gyr (rms), and a possible age$-$metallicity 
relation. No correlation between age and Galactocentric distance has been
conclusively demonstrated.

In this paper, we present an analysis of a uniform set of
HST/ACS data for 64 GGCs\footnote{\citet{Sar07}
presented observations of 65 GCs, but Pal 2 has been left out of
this study due to its high differential
reddening}.  This dataset, previously introduced
by \citet{Sar07}, has produced well-defined CMDs for
all of the target clusters from the tip of the red giant branch
(RGB) to at least $\sim$6.5 magnitudes below the MSTO \citep{A08}.
The determination of GGC ages was a principle goal of the HST/ACS survey
and exposures times were defined to maximize the S/N ratio
at the MSTO.  This makes the survey an excellent resource for the evaluation
of relative ages.

Due to the difficulty in determining the position of the MSTO\footnote{
The MSTO point is the current position in the CMD where stars in the cluster start 
burning hydrogen in a shell and are leaving the main$-$sequence. 
The empirical MSTO is then the bluest point on the main$-$sequence$-$to$-$subgiant$-$branch 
portion of the cluster's CMD.}
to the high precision required for age studies,
many different methods have been developed in previous relative ages works 
\citep[see][for a complete discussion]{MW06}. The two most used relative age
indicators are those based on the color difference between the
MSTO and the RGB at a given magnitude level (horizontal method), and
the magnitude difference between the MSTO and the zero$-$age HB\footnote{
The zero$-$age HB is the position in the CMD where stars in the cluster start burning helium
into carbon and oxygen in a helium rich core and hydrogen into helium in a shell.}
(vertical method). 
\citet{VBS90} were pioneers in measuring GGC relative ages making use of the horizontal method. 
They essentially derived relative ages comparing the color difference between the MSTO
and the location of a well defined point in the lower RGB, located 2.5 magnitudes brighter than a 
point in the upper main$-$sequence (MS) that is 0.05 magnitudes redder than the MSTO. 
However, horizontal method results are sensitive to
the assumed value of the mixing length parameter and 
to the color$-$T$_{\rm eff}$ transformation of the theoretical models
into the observational plane. As a result, relative ages determined using the
horizontal method are model dependent.  Potentially, more
reliable age indicators are those related to the brightness of the
MSTO, in particular the vertical method. 
First surveys of GGC relative ages using the vertical method were carried out by
\citet{G85} and \citet{P87}. They measured relative ages from a sample of 26 and 41 GGCs,
respectively, collected from the available literature. Later on, \citet{SK89} used the brightness
difference between the HB and the MSTO to estimate the ages of 31 GGCs.
Nevertheless, accurate
determination of the zero$-$age HB is delicate, in particular for the
clusters at the extreme boundaries of the metallicity distribution,
which generally have a red (or blue) HB and no RR Lyrae
stars. Moreover, the dependence of the zero$-$age HB luminosity on the
cluster metallicity is still controversial.

In this paper, we use an alternative approach. Our data reaches 
at least $\sim$6.5 magnitudes below the MSTO, providing a very well 
defined MS. This allows, for the first time,
the use of relative MS fitting for a large number of clusters.  
As it will be shown later, the relative MS fitting procedure used in this study does essentially
the following. The MS of two clusters, with similar metallicities, 
are superimposed in the CMD by shifting one of the cluster's mean ridge line
in both color and magnitude  (explained 
in detail in section~\ref{MRLandMSF}). This way, any differences between the two cluster's 
distance and reddening are compensated, and an intrinsic MSTO magnitude difference
is obtained. This relative MSTO brightness is then used to derive relative ages.
This allows us to use what is, in essence, an improved version of the 
traditional vertical method that substitutes the well-defined MS for the contentious
zero$-$age HB.

We also take the additional step of comparing our observational measures to
updated theoretical models. We have primarily used the most recent state$-$of$-$art models of
\citet{D07} (hereafter D07), transformed to the observational plane with
an empirical Color$-$T$_{\rm eff}$ transformation.  However, we also compare our results to the models of \citet{P04} (hereafter P04),
\citet{B94} (hereafter B94) and \citet{G00} (hereafter G00).  We find that the various
theoretical models agree closely on the MSTO absolute magnitude and color, resulting
in negligible model dependence.

This paper is organized as follows. Section~\ref{obs} describes the
observations and data reduction pipeline, and 
Section~\ref{metScale} presents the globular cluster database. The mean
ridge line fitting and MS$-$fitting procedures used in this
study are described in Section~\ref{MRLandMSF}. The analysis of the
uncertainties is presented in Section~\ref{errorDet}. Finally,
relative age results and a discussion are presented in
Sections~\ref{results} and \ref{discussion}, respectively. The paper
ends with a summary of conclusions 
(Section~\ref{conclusions}).

\section{Observations and data reduction}\label{obs}

Observations were performed with the ACS/WFC instrument on board 
the $HST$. A total of 65 GGCs were observed in the $F606W$ ($\sim V$) 
and $F814W$ ($\sim I$) filters.  Due to its high differential reddening, 
Pal 2 is not considered here.

Details on the observations, data reduction and photometry are given in 
\citet{Sar07} and \citet{A08}. For most globular clusters, the observation and 
data reduction pipeline produces $\sim$12 magnitudes of precise photometry, extending
from nearly the tip of the RGB to several magnitudes below the MSTO.  In general, 
the photometry reaches a depth corresponding to an approximate stellar mass of $\sim$ 0.2 $m_\odot$. 

This uniform and deep photometry offers a database of unprecedented
quality, opening a window for new data analysis approaches. In particular, 
cluster MSs are so well defined that it is possible, for the first time,
to apply relative MS$-$fitting techniques to a large number of clusters within the same database.

\section{The database}\label{metScale}

The position of the MSTO depends not only upon age but also upon metallicity. This
effect requires the use of a large and homogeneous metallicity database in order to
determine accurate relative ages.
To account for this, we have adopted the metallicities listed in the
\citet{R97} catalog\footnote{\citet{R97} presented a large homogeneous metallicity database 
(over 71 GGCs) based on calcium index (Ca II Triplet) measurements.} over both the \citet{CG97}
(hereafter CG) and the \citet{ZW84} (hereafter ZW) metallicity scales. 
This catalog is maybe not as precise as a true compilation of metallicities based on high dispersion
spectra, but it has the the advantage of being homogeneous. For clusters not listed in \citet{R97}, the original 
metallicity listed in the ZW catalog has been transformed to the CG metallicity scale using
the following equation \citep{C01}:

\begin{equation}
[Fe/H]_{\rm CG} = 0.61 + 3.04[Fe/H]_{\rm ZW} + 1.981[Fe/H]_{\rm ZW}^{2} + 0.532[Fe/H]_{\rm ZW}^{3}
\end{equation}

Lyng\aa\  7, E3, and Pal 1 metallicities are not listed in the previous catalogs. Lyng\aa 7 
and E3 metallicities in the CG scale have been calculated from the ZW metallicities
of \citet{TF95} and \citet{H96}, respectively. For Pal 1, we used the measure of
\citet{R98}.

Table~\ref{GCgroups} presents the list of 
our target GGCs, grouped by metallicity.  Column 3 gives the cluster identity while columns 4 and 5 present the adopted 
metallicities in the ZW and CG scales, respectively. Column 6 lists the Galactocentric 
distance ($r_{\rm GC}$, in kpc), taken from \citet{H96}. 

\section{Mean ridge lines and main sequence fitting method}\label{MRLandMSF}

In this section, we describe the general properties of the cluster
mean ridge line fitting procedure.  We then use the derived ridge lines to
perform relative MS$-$fitting between clusters.

\subsection{Mean ridge lines and main sequence turn-off apparent magnitude}\label{MRLdet}

In order to determine accurate relative ages for each cluster, a
precise determination of the MS and RGB mean ridge line is
required. With this aim, we have created a new software package for the ACS program.
It is worth mentioning that mean ridge lines are determined 
in the observational plane ($F606W$-$F814W$, $F814W$) instead of ($F606W$-$F814W$, $F606W$),
as the SGB appears more vertical in the CMD and the following procedure 
produces more accurate results. For the rest of the analysis, the ($F606W$-$F814W$, $F606W$) plane
is used.

\begin{figure}
\epsscale{1.00}
\plotone{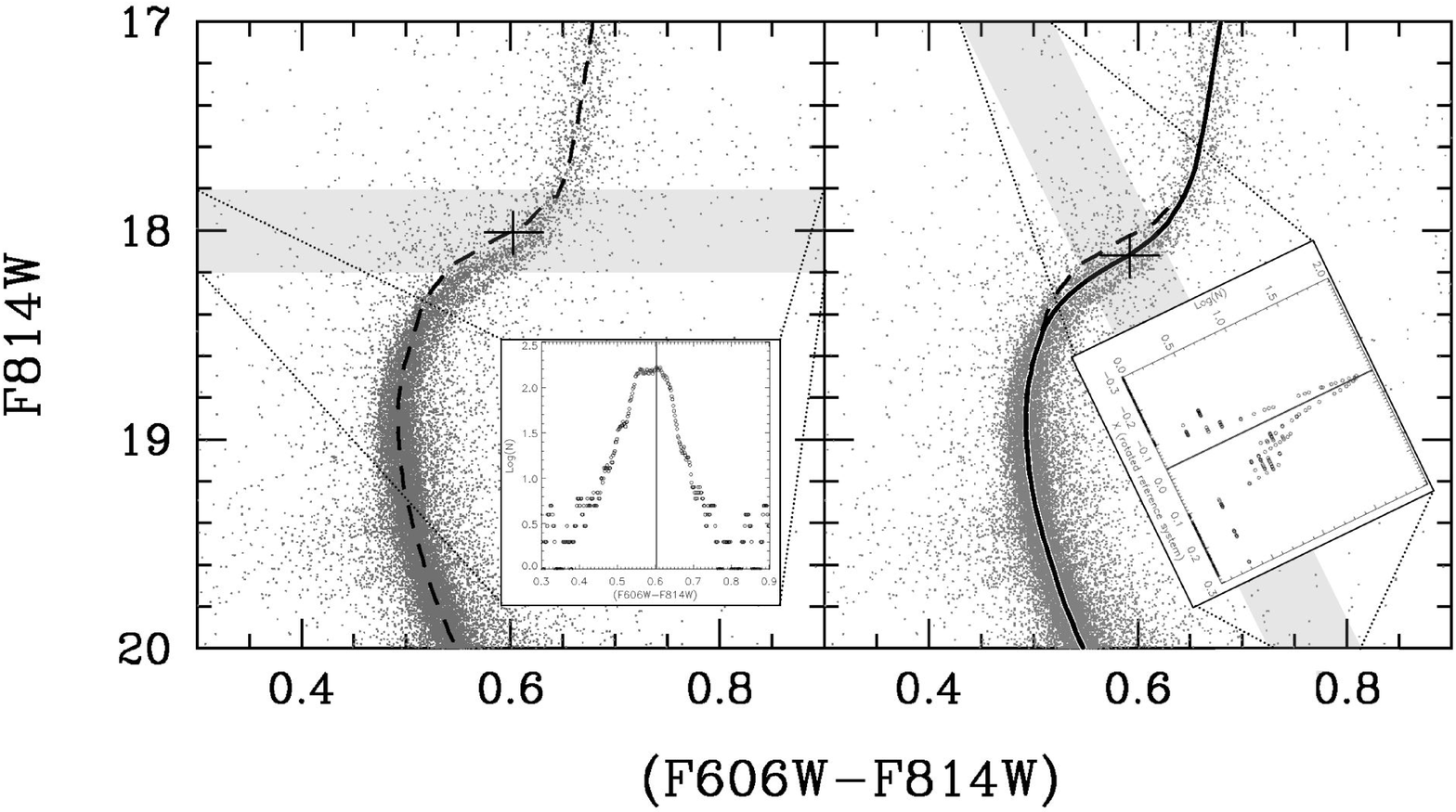}
\caption{Mean ridge line fitting procedure using {\it rotated} histograms. This shows the analysis of NGC 1851. 
In short, the mean ridge line is determined in two steps. First, color histograms are computed for stars grouped 
in 0.4 magnitude ($F814W$) wide bins (left panel). A preliminary mean ridge line is then derived with 
$(F606W-F814W)$ colors set at each histogram's maximum, and the $F814W$ magnitude of the bin center.
Second, we use $\it rotated$ histograms, perpendicular to the preliminary mean ridge line, to derive 
the final one (right panel). See text for details.
\label{MRLfittingProc}}
\end{figure}

For each globular
cluster, the mean ridge line is determined in two steps 
(Figure~\ref{MRLfittingProc}).  First,
color histograms are computed for stars grouped in 0.4 magnitude
($F814W$) wide bins. The value of 0.4 magnitudes was determined 
experimentally as a compromise between two effects -- the low
resolution produced by larger bins and the noisy ridge line produced
by smaller bins.  Color histograms are constructed along the RGB and
MS with a step of 0.04 magnitudes (a tenth of the bin size) in F814W. 
That is, a moving bin of fixed width (0.4 magnitudes),
with successive steps of 0.04 magnitudes, is used. A preliminary mean ridge line is then derived with 
$(F606W-F814W)$ colors set at each histogram's maximum, and the $F814W$ magnitude
of the bin center.

\begin{figure}
\epsscale{1.00}
\plotone{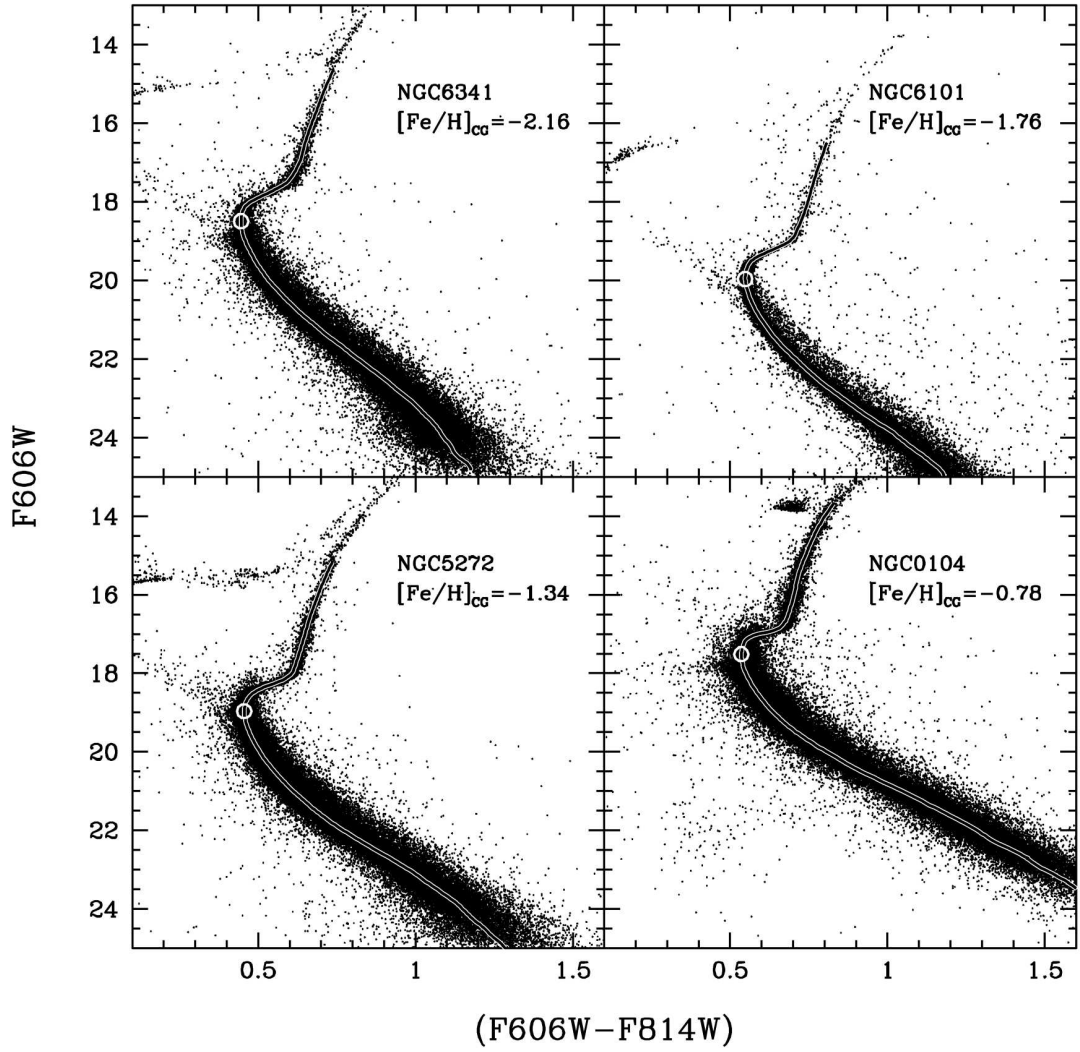}
\caption{Examples of mean ridge line fitting in four selected clusters covering the metallicity range. The MSTO position is marked with an open circle. \label{MRLexample}}
\end{figure}

Once the preliminary mean ridge line has been determined, the fitting
process is iterated three more times.  However, we now use $\it rotated$ histograms.
For each $F814W$ value, the tangential angle of the mean ridge line is
determined. This angle is used to rotate the reference
system on which the histogram is computed.  The X$-$axis
of the new reference system is perpendicular to the preliminary mean
ridge line at the considered $F814W$ magnitude. The 
histogram is then computed perpendicular to the preliminary mean ridge line
and considering stars in the new reference system's Y$-$axis interval [-0.04, 0.04].
The histogram's maximum is determined, defining a point of the
new mean ridge line in the rotated reference system, ($\it max$
,0). Note that this point shows the correction to be applied to the
preliminary mean ridge line. The coordinates of this point are
$\it de-rotated$ and its location in the original reference system,
the CMD, is calculated. The process is repeated for all the $F814W$ values to derive
the revised mean ridge line, which is then smoothed to derive the final mean ridge line.
Figure \ref{MRLexample} shows 
some examples of mean ridge line fitting results. Four selected clusters 
covering the metallicity range are shown. 

For clusters with multiple stellar populations,
such as NGC 1851  \citep{M07}, NGC 2808 \citep{P07}, M54  \citep{S07}
and $\omega$Cen \citep{L99, P00, B04}, the mean ridge line fitting procedure
fits  the dominant population (see Figure~\ref{MRLfittingProc}).

The empirical MSTO is the bluest point on the mean ridge line. 
It is worth mentioning that the final mean ridge line consists of a discrete number of 
points, it is not a continuous line. For this reason, in order to determine the MSTO 
position, a continuous spline has been fitted to the 
main$-$sequence$-$to$-$subgiant$-$branch portion of the final mean ridge line. 
Its bluest point is then adopted as the empirical MSTO.  
As an example, the position of the MSTO is marked with an open circle in Figure~\ref{MRLexample}.
In order to increase the precision of the MSTO position, the mean ridge lines 
have been determined ten times for each cluster, with the starting point of the histograms
offset in steps of 0.004 magnitudes. The MSTO color 
and apparent magnitude have been determined as the mean of the ten 
different measurements.

This method provides a consistently defined mean ridge line, minimizing the effects 
of the binary population and differential reddening (see section \ref{errorDet}). Additionally, the rotated
coordinate system ensures that the best fit is derived
around the steeply curved MSTO and sub giant branch region of the CMD. For high metallicity clusters, 
in particular, the sub giant branch is nearly horizontal 
and the more commonly used mean ridge line determination method (drawing 
an approximate ridge line, rendering the sequence to a nearly vertical line and
then taking histograms in small ranges of magnitude) lacks the resolution
around the sub giant branch needed for precise age measurement.

\subsection{Main sequence fitting method}\label{MSFmethod}

\begin{figure}
\epsscale{1.00}
\plotone{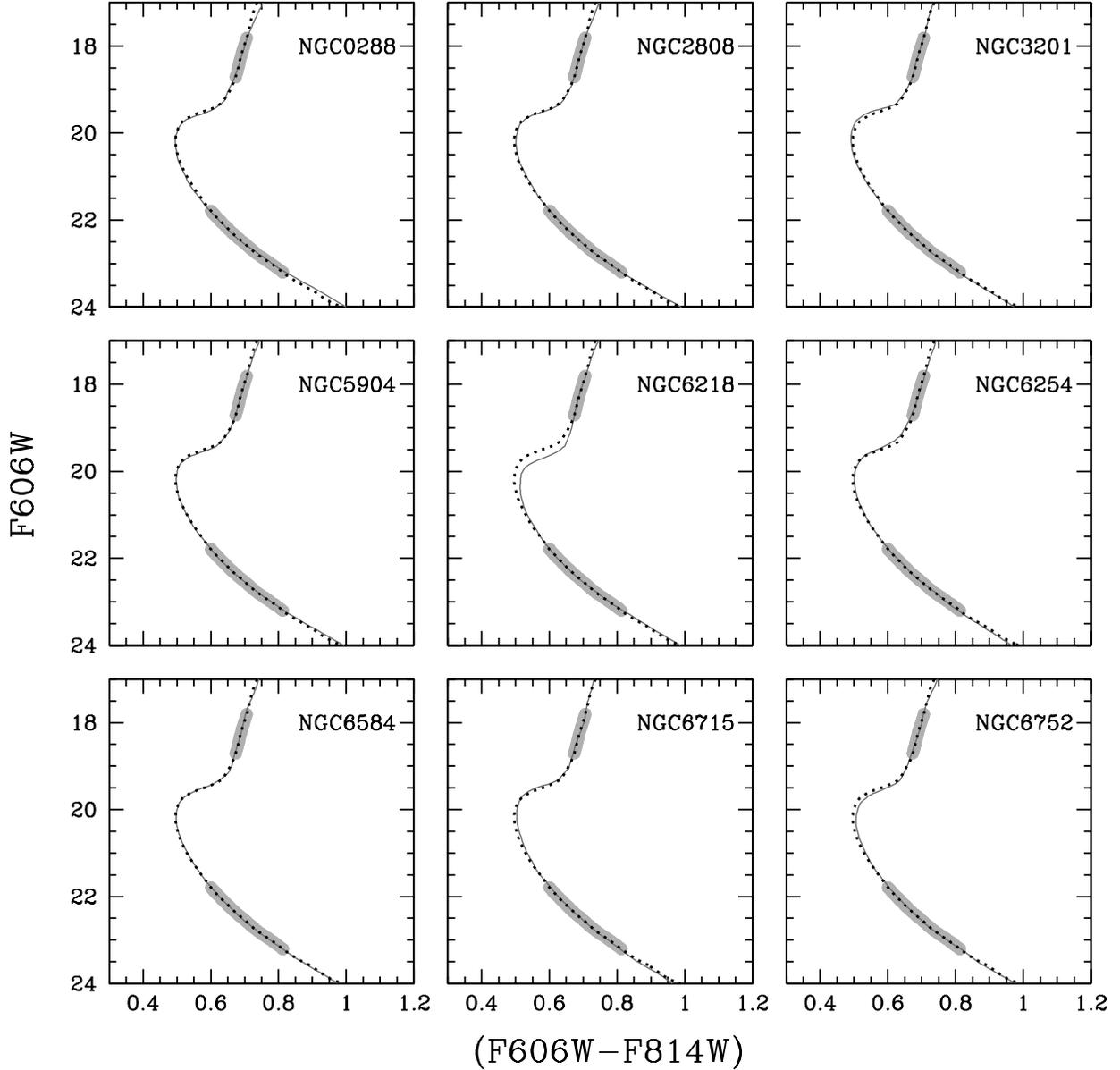}
\caption{Examples of MS$-$fitting for the $-1.3\le[Fe/H]_{\rm CG}<-1.1$ metallicity group. 
The reference cluster is NGC 6981 (dashed line). Each cluster MRL (solid line) has been fitted 
to the reference cluster in the magnitude intervals 
[$(M_{\rm F606W}^{\rm TO}-2.5)<M_{\rm F606W}<(M_{\rm F606W}^{\rm TO}-1.5)$] and 
[$(M_{\rm F606W}^{\rm TO}+1.5)<M_{\rm F606W}<(M_{\rm F606W}^{\rm TO}+3.0)$] 
(shaded regions). \label{MSFexample}}
\end{figure}

The CMD location of the faint MS of a GGC is independent 
of its age, but highly dependent upon its metallicity. For this reason, we have
divided the HST/ACS sample into six metallicity groups (see Table~\ref{GCgroups}) using the following
divisions: [Fe/H] = -0.3, -0.8, --1.1, -1.3, -1.5, -1,8 and -2.3.
For each group, a reference cluster has been chosen: NGC 6304, NGC 6723,
NGC 6981, NGC 6681, NGC 6101 and NGC4590, respectively. Reference clusters 
have been selected to have low differential reddening, low field star
contamination, and a well defined MS. Using the derived mean ridge
lines, we performed MS$-$ and RGB$-$fitting between each cluster in each metallicity group
and the corresponding reference cluster. Figure~\ref{MSFexample} shows examples of the fitting, in which
the mean ridge lines of clusters with $-1.3 \le [Fe/H]_{\rm CG} < -1.1$
(solid lines) have been shifted in both magnitude and color to fit the
reference cluster, NGC 6981 (dashed line). The fitting was performed in a least squares
fashion and taking into account two CMD regions that have little dependence upon cluster age.  The
two intervals are [$(M_{\rm F606W}^{\rm TO}-2.5)<M_{\rm F606W}<(M_{\rm F606W}^{\rm TO}-1.5)$] and
[$(M_{\rm F606W}^{\rm TO}+1.5)<M_{\rm F606W}<(M_{\rm F606W}^{\rm TO}+3.0)$],
$M_{\rm F606W}^{\rm TO}$ being the magnitude of the reference cluster's MSTO.  
These two intervals have been defined making use of D07 theoretical isochrones. 
In particular, D07 isochrones with similar metallicity and different ages were
superimposed on the same CMD. Based on visual inspection, these two particular 
regions were found to have little dependence upon cluster age, and were adopted
as the optimum intervals for the MS fitting procedure.
These regions
are shaded in Figure \ref{MSFexample}. This fitting procedure produces unequivocal results,
with no degeneracy between distance modulus and reddening.

The MS$-$fitting provides relative MSTO magnitudes 
for clusters inside each metallicity group. In order to calculate relative MSTOs between 
two consecutive metallicity groups, we applied the MS$-$fitting technique between the 
corresponding reference clusters. This was repeated for every two adjacent metallicity groups, producing
a unified sample of relative MSTO magnitudes for all the analyzed clusters.  

In order to compare relative MSTO measures to model predictions, 
we had to calculate the {\it absolute} magnitudes of the turnoffs.
As the relative MSTO magnitudes are known, the MSTO absolute magnitude is only needed for
one cluster. We chose the well-measured NGC 6752 cluster, adopting a subdwarf-based distance
modulus of $(m-M)_V=13.24\pm0.08$ from \citet{G03}.  We calculate an MSTO magnitude
of $M_{F606W}^{TO}=4.02\pm0.08$ for NGC 6752.

We performed similar calculations using the NGC 0104 and NGC 6397 distance moduli
listed in \citet{G03} of $(m-M)_V=13.50\pm0.08$ and $12.58\pm0.08$ , respectively.
We calculate MSTO absolute magnitudes of $M_{F606W}^{TO}=4.03\pm0.08$ and
$M_{F606W}^{TO}=3.77\pm0.08$ for NGC 0104 and NGC 6397, respectively.
As the relative MSTOs are known, NGC 6752's MSTO absolute magnitude can be also
derived independently from NGC 0104 and NGC 6397 ($3.72\pm0.08$
and $3.86\pm0.08$, respectively). The mean of the three NGC 6752 MSTO absolute magnitude 
determinations is $M_{F606W}^{TO}=3.87\pm0.15$ and we adopt this as the final MSTO zeropoint.

We note that any error in the MSTO magnitude zeropoint 
would affect {\it absolute} age determination, but not {\it relative} age measurements.
For this reason, the zeropoint uncertainty is not
taken into consideration during relative age determination. 

Using the adopted zeropoint, we calculate MSTO absolute magnitudes for
our entire sample, which are listed in column 4 of Table~\ref{GCages}.
The listed uncertainties are determined in next section. 

\section{Main sequence turn off uncertainties}\label{errorDet}

Several effects must be taken into account when deriving the uncertainty
associated with the measured MSTO magnitudes ($M_{\rm F606W}^{\rm TO}$).
Among these, we address the cluster differential reddening, the binary
fraction and total number of cluster stars. The
total number of stars affects the MSTO magnitude
measurements in two ways. The first one is related to the total number of stars close
to the MSTO that can be used to measure its magnitude: the smaller this number,
the more uncertain the MSTO magnitude will be. On the other hand, a large
number of cluster stars implies more severe crowding, and therefore less
accurate magnitude measurement. To these uncertainties, we need to add
the uncertainty from the MS$-$fitting and the mean ridge line
determination procedure itself. The final uncertainty associated to the MSTO
magnitude, $\sigma^{2}(M_{\rm F606W}^{\rm TO})$, is:

\begin{equation}\label{sigmaeq}
\sigma^{2}(M_{\rm F606W}^{\rm TO}) = \sigma^{2}_{\rm RED} + \sigma^{2}_{\rm BIN} + \sigma^{2}_{\rm NUM}  +
\sigma^{2}_{\rm CROW}+ \sigma^{2}_{\rm MRL} + \sigma^{2}_{\rm MSF}
\end{equation}

\noindent where $\sigma^{2}_{\rm RED}$ is the variance due to the
differential reddening, $\sigma^{2}_{\rm BIN}$ due to the binary star
population, $\sigma^{2}_{\rm NUM}$ due to the number of cluster stars,
$\sigma^{2}_{\rm CROW}$ due to the crowding, 
$\sigma^{2}_{\rm MRL}$ is the MSTO's magitude variance
from the mean ridge line determination, and
$\sigma^{2}_{\rm MSF}$ from the MS$-$fitting method. 
It is worth noting that field star contamination
does not affect the MSTO determination. Our fields cover the central
region in each cluster and the ratio of cluster stars to field stars
is very high.  Moreover, as noted above, the mean ridge line fitting method
fits the dominant population, ignoring both field stars and blue stragglers.

As it has been described in Section~\ref{MRLdet}, each MSTO's
magnitude has been determined as the mean of the MSTO's magnitude obtained from ten 
different mean ridge line fittings. Here we assume that the variance associated 
with the mean ridge line determination, $\sigma^{2}_{\rm MRL}$, is the variance 
of the previous ten measurements. Typical values of $\sigma_{\rm MRL}$ 
are in the range $\sim$0.01-0.04 mag.

With the aim of performing an external cross-check for the $\sigma^{2}_{\rm MRL}$, 
the MSTO's magnitude variance from to the mean ridge line determination, we have considered four clusters having low 
reddening and a large number of stars (NGC 0104, NGC 5024, NGC 5272 and NGC7089).
These clusters' CMDs have been randomly divided into 5 partial CMDs. A mean ridge line has been
fitted to each partial CMD and its MSTO's magnitude has been measured. Finally, 
$\sigma/\sqrt{5-1}$ for these five MSTO's measurements has 
been computed for each cluster, 
where $\sigma$ is the standard deviation of the five measurements. 

Results are shown in 
Table~\ref{sigmaMRLtest}. Column 1 shows the cluster's name, column 2 lists $\sigma_{\rm MRL}$
and column 3 lists $\sigma_{cross-check}=\sigma/\sqrt{5-1}$ derived in this external cross-check. It can be seen that 
the uncertainties obtained here are equivalent to the derived $\sigma_{\rm MRL}$. This test
indicates that the adopted $\sigma_{\rm MRL}$ is a good estimation.

For $\sigma^{2}_{\rm MSF}$, we adopted the uncertainty calculated by the least squares 
main sequence fitting procedure. Typical values of $\sigma_{\rm MSF}$ 
are in the interval $\sim$0.01-0.06 mag.

To estimate the uncertainties induced by differential reddening, binary star
population, and total number of cluster stars, as well as by differences between 
the measured MSTO and the actual GGC's MSTO, we generated over 250 synthetic
CMDs using the IAC$-$star \citep{AG04} synthetic CMD program.
Stellar populations are calculated on a star-by-star basis. $L$, $T_{\rm eff}$, and
$g$ of each star are computed by direct bilogarithmic interpolation in
both the metallicity and age grids of a library of stellar evolution tracks,
metallicity and age following continuous distributions. All the stellar evolution
phases are covered accurately, and mass loss is computed during the
RGB and the asymptotic giant branch phases
\citep[see][for details]{AG04}. A set of
different star formation rates, initial mass functions and metallicity laws
are allowed, and binary star content can be computed.
The result is a synthetic CMD with a smooth realistic stellar distribution.

Synthetic CMDs were generated with an age of 13 Gyr, $[M/H]
= -1.2$, differential reddening $E(F606W-F814W)$ values between 0.0 and
0.2, binary fractions (with $0.5<m_2/m_1<1.0$) between
0.0 and 0.7, and a number of stars brighter than $M_{\rm F606W} = 6$ 
(approximately two mag. below the MSTO) between
1,000 and 50,000. The P04 stellar evolution library was
used to generate the stellar libraries. For each combination of differential reddening,
binary fraction and number of stars, 10 different CMDs were generated.
Observational errors were simulated in all the
synthetic CMDs by applying
a gaussian dispersion to the synthetic CMD, with $\sigma$ taken
from the dispersion of the observed CMD as a function of the magnitude.

\begin{figure}
\epsscale{1.00}
\plotone{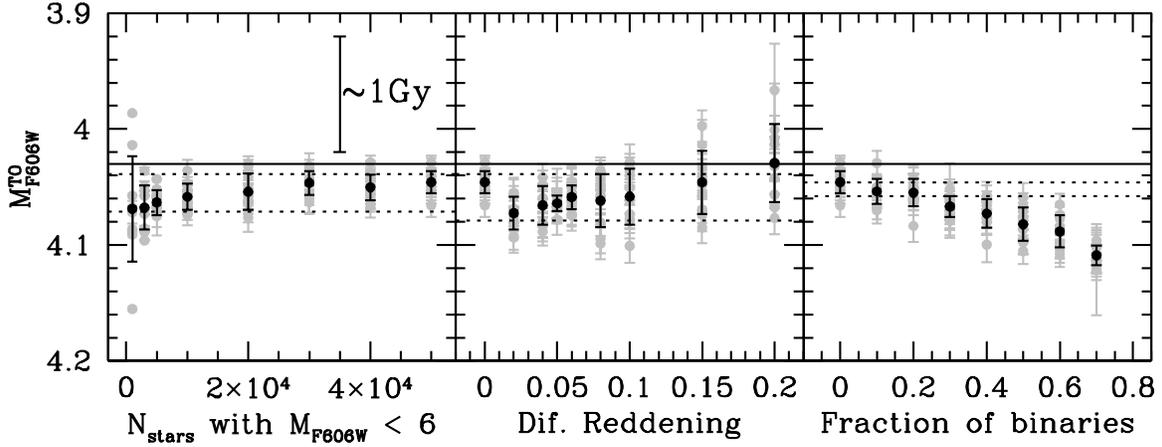}
\caption{The right panel shows $M^{\rm TO}_{\rm F606W}$ measurements over synthetic
CMDs with 50,000 stars brighter than $M_{\rm F606W} = 6$, no differential
reddening, and varying the fraction of binary stars. Grey points show
the MSTO measurements for the 10 different CMDs generated in each case,
and black points show the average and standard deviation of each of
the 10 measurement sets. Solid line represents the MSTO's magnitude measured directly on the
isochrone. The Central panel shows the same as left panel,
but now the fraction of binary stars is 0.0 and the maximum differential
reddening is varying between 0.0 and 0.2. The left panel shows the same, but
now the fraction of binary stars is 0.0, the differential reddening
is 0.0 and the number of stars brighter then $M_{\rm F606W} = 6$ is
varying between 1,000 to 50,000. Dashed lines represent the adopted $\pm
\sigma$ in each case. See text for details. \label{TOerror}}
\end{figure}

Using the same mean ridge line fitting method used for the observational data,
MSTO's magnitudes were measured for the synthetic CMD.  The results
are shown in Figure~\ref{TOerror}. Grey points show 
the MSTO measurements from the 10 different CMDs generated for each experiment, 
black points show the mean, and the error bars correspond to the standard 
deviation of the 10 measurements. Solid line represents the input MSTO 
magnitude, that is, the magnitude of the MSTO obtained directly from the 
input isochrone. 
It can be seen that a systematic effect on the measured MSTO's magnitudes exists. 
Output MSTOs (measured) are systematically $\sim$0.02 mag. fainter than input ones. This would have an effect on
the measured absolute ages, but as this work is focused on relative ages, the variance 
of the difference between the input and output MSTO's magnitudes, $\sigma^{2}_{\rm IO}$, will be 
also considered. 

The left panel of Figure~\ref{TOerror} shows the $M^{\rm TO}_{\rm F606W}$ measurements 
for the synthetic CMDs with no differential reddening, no binary stars and varying the number of stars 
brighter than $M_{\rm F606W} = 6$ from 1,000 to 50,000. 
With the aim of determining $\sigma^{2}_{\rm NUM}$, a gaussian has been generated 
for each black point, being the mean and the sigma of the gaussian equal to the value of the 
corresponding point and its sigma, respectively. Finally, all gaussians have been added
together and the variance of this sum of gaussians has been adopted as $\sigma^{2}_{\rm NUM}$.
Since most of the target clusters have more than 1,000 stars brighter than $F606W=6$,
we adopt $\sigma_{\rm NUM}=1.6\times10^{-2}$ mag. (dotted lines). 
Estimated this way, $\sigma^{2}_{\rm NUM}$ also includes the $\sigma^{2}_{\rm IO}$ contribution.

The central panel of Figure~\ref{TOerror} shows the effect of differential reddening upon the
estimated $M^{\rm TO}_{\rm F606W}$ value. For a synthetic CMD, the reddening has been 
simulated by shifting every star along the reddening vector a quantity between
zero and a maximum differential reddening value, $E(F606W-F814W)_{max}$. 
We generated a set of nine synthetic CMDs with 50,000 stars brighter 
than $F606W=6$ and different values of $E(F606W-F814W)_{max}$
from 0.0 to 0.2 magnitudes. 
It is worth noting that the black points' variance also includes the 
contribution from $\sigma^{2}_{\rm IO}$. In order to avoid duplicity, the contribution 
of $\sigma^{2}_{\rm IO}$ must be eliminated. As these synthetic diagrams have 50,000 stars
brighter than $F606W=6$, their $\sigma^{2}_{\rm IO}$ is equal to the variance
of the point in the left panel corresponding to a number of stars brighter than 
$F606W=6$ equal to 50,000, that is $\sigma_{\rm IO}= 9.7\times10^{-3}$ mag. 
In order to compute the effect of differential reddening upon the
estimated $M^{\rm TO}_{\rm F606W}$ value, $\sigma^{2}_{\rm IO}$ has been
quadratically subtracted to each point's variance in the central panel of Figure~\ref{TOerror}. 
The resulting variance accounts for the differential reddening effect only. Then a gaussian 
has been generated for each black point, being the mean and the sigma of the gaussian 
equal to the value of the corresponding point and the obtained quadratically subtracted 
sigma, respectively. Finally, all gaussians have been added
together and the variance of this sum of gaussians has been adopted as 
$\sigma^{2}_{\rm RED}$. As we expect most of the target clusters to have differential
reddening lower than 0.15 magnitudes, we adopt $\sigma_{\rm RED} = 2.0\times10^{-2}$ mag.

The right panel of Figure~\ref{TOerror} shows the $M^{\rm TO}_{\rm F606W}$ 
measurements of the synthetic CMDs with 50,000 stars
brighter than $M_{\rm F606W} = 6$, no differential reddening, and varying
fractions of binary stars. It can be seen that the fraction of binary stars affects
$M^{\rm TO}_{\rm F606W}$, making the measured MSTO fainter for clusters
with a higher binary fractions. Following the same procedure
as before, 
$\sigma^{2}_{\rm IO}$ has been quadratically subtracted to each point's variance in 
the right panel of Figure~\ref{TOerror}. Then a gaussian 
has been generated for each black point, being the mean and the sigma of the gaussian 
equal to the value of the corresponding point and the
sigma obtained by subtracting $\sigma^{2}_{\rm IO}$ to the rms of the point, 
respectively. Finally, all gaussians have been added
together and the variance of this sum of gaussians has been adopted as 
$\sigma^{2}_{\rm BIN}$. In a parallel work, we are measuring
the fraction of binary stars in each cluster in our database \citep{M08}.  For most
clusters, it is smaller than 20\%. Accordingly, we consider the fraction of binary stars 
interval [0.0, 0.2] which results in $\sigma_{\rm BIN} = 5.9\times10^{-3}$ mag.

To estimate the effect of cluster crowding on the measured MSTO magnitude, we used the artificial$-$star tests
described in \citet{A08}. In principle, crowding would produce some systematic shift between output 
and input magnitudes of artificial stars. We are interested in evaluating if these shifts significantly change 
as a function of crowding. Since different clusters have different crowding levels, determining the artificial 
star magnitude shifts in several of them will provide the information we need. We will use the NGC 0104, 
NGC 1851, NGC 2808, NGC 5139, NGC 5286, NGC 6171 and E3 clusters. This sample contains both
one of the most and one of the least crowded clusters, namely NGC 2808 and E3, respectively.

Each cluster has associated a table containing input and output positions and magnitudes of the artificial stars.
For each cluster, we selected artificial stars with input $F814W$ ($I$) magnitudes within $\pm$0.25 mag of 
the MSTO. For most clusters, $\sim 4,000$ artificial stars fulfill this condition. For each of them, we calculated 
the quantity $\delta I = I_{out} - I_{in}$,
where $I_{out}$ and $I_{in}$ are the recovered and input magnitudes, respectively. To have a representative 
estimate of $\delta I$ and also of $\sigma(\delta I)$ for each cluster, we have divided the artificial stars in 10 
subsamples and calculated med$(\delta I)$ for each of them. The average of medians, $<{\rm med}(\delta I)>$, 
and the mean square root, $\sigma[{\rm med}(\delta I)]$, of the 10 med($\delta I$) values have been computed 
for each cluster. Standard errors $\sigma_n=\sigma[{\rm med}(\delta I)]/\sqrt{(n-1)}$, with $n=10$, are obtained. 
These can be considered as an estimate of the internal errors of ${\rm med}(\delta I)$ of each cluster. Average 
values $<{\rm med}(\delta I)>$ run from $3.0\times 10^{-4}$ mag for E3, to $-4.5\times 10^{-3}$ mag for 
NGC2808, while $\sigma_n$ is always of the order of $1.0\times10^{-4}$ mag, indicating that 
the ${\rm med}(\delta I)$ are significant. In summary, we can say that $\sim 5 \times 10^{-3}$ mag is the 
maximum shift introduced by crowding in our MSTO measurements. Finally, we can estimate the uncertainty 
produced by crowding that we were looking for,  $\sigma_{\rm CROW}$, as the root mean square of 
the $<{\rm med}(\delta I)>$ values of each cluster. This results in $\sigma_{\rm CROW}=1.5\times10^{-3}$ mag.

Table~\ref{stddevs} lists a summary of the different contributions to $M^{\rm TO}_{\rm F606W}$ uncertainties. 
The MSTO's magnitude variances produced by the cluster number of stars, differential reddening, 
binary stars content and crowding, evaluated in this section using simulations, have been quadratically added
and the resulting sigma (0.026 mag.) is also shown in Table~\ref{stddevs}. 
Finally, $M^{\rm TO}_{\rm F606W}$ uncertainty has been evaluated for each
cluster adding quadratically 0.026 mag. to the obtained $\sigma^{2}_{\rm MRL}$ 
and $\sigma^{2}_{\rm MSF}$ (equation \ref{sigmaeq}), and the results are listed in 
Table~\ref{GCages} (column 4). It is worth mentioning that the total uncertainties tend to increase both
at low and high metallicities. This is due to the error propagation in the $\sigma^{2}_{\rm MSF}$
contribution, as we are using the intermediate metallicity cluster NGC 6752 for the MSTO absolute 
magnitude zeropoint.

\section{Results}\label{results}

In this section, we present the relative ages derived for the sample of 64 GGCs. 
We compare the derived ages to the theoretical isochrones of D07 using both the ZW and CG abundance scales.
In the second part of 
this section, we test the MS$-$fitting procedure and check the self$-$consistency of the relative age determination.
The relative age results are then compared with those 
derived using the stellar evolution libraries of P04, B94, and G00. 
Finally, we present a comparison of our results with previous work.

\subsection{Relative ages}\label{relAges}

\begin{figure}
\epsscale{1.00}
\plotone{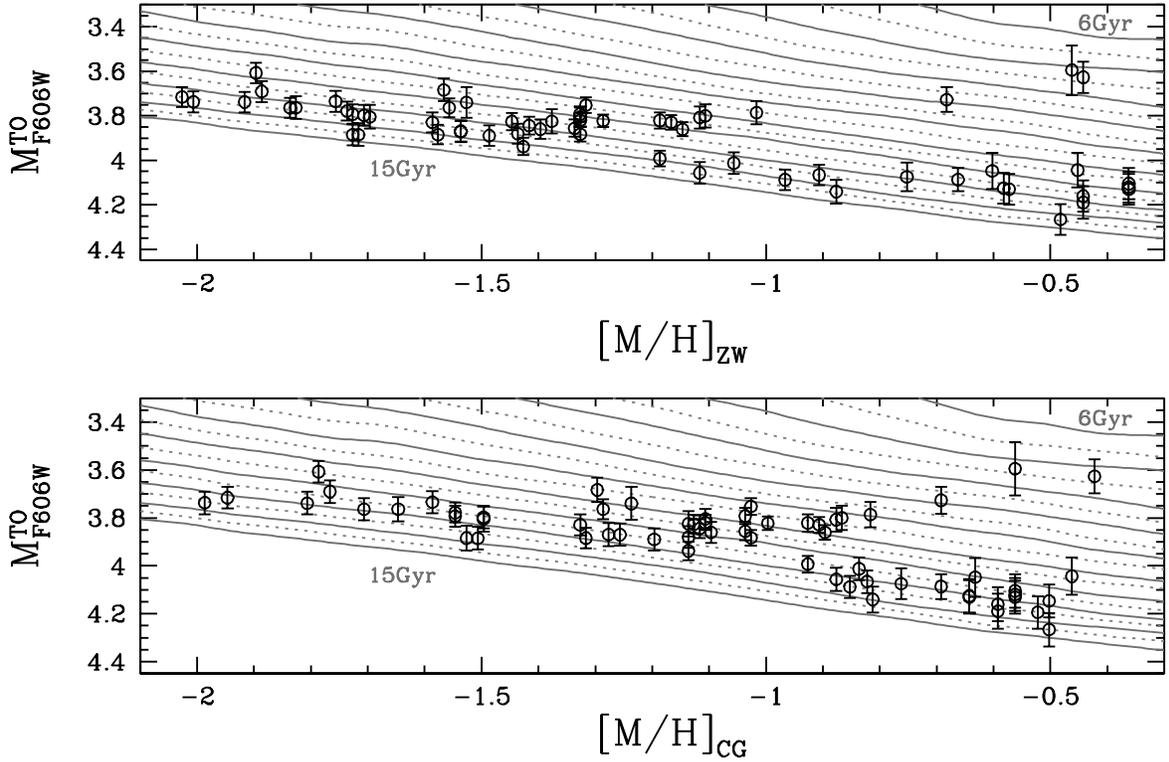}
\caption{Theoretical TO absolute magnitudes from the D07 stellar evolution
library. Lines represent $M^{\rm TO}_{\rm F606W}$ for different ages as a
function of global metallicity, in 1 Gyr (solid lines) and 0.5 Gyr
(dashed lines) age steps. The lower isochrone corresponds to an age of
15 Gyr. Points represent the measured MSTO magnitudes. 
The upper panel shows the $M^{\rm TO}_{\rm F606W}$ magnitude as a function of
the metallicity in the ZW scale; the lower panel gives
$M^{\rm TO}_{\rm F606W}$ as a function of the metallicity in the CG scale.
\label{TOModelsZW}}
\end{figure}

Relative ages were calculated using a stellar evolution 
library (D07) to calculate the theoretical $M^{\rm TO}_{\rm F606W}$ values for different ages and [M/H]. 
Figure~\ref{TOModelsZW} shows the $M^{\rm TO}_{\rm F606W}$ resulting from the D07 models as a 
function of [M/H], using both ZW and CG metallicity scales. 
Lines represent the D07 model MSTO magnitudes in steps of 1 Gyr (solid lines) and 
0.5 Gyr (dotted lines). The curves are interpolated using a spline curve so that we can easily
estimate $M^{\rm TO}_{\rm F606W}$ = f([M/H], age). We overplot the turnoff magnitudes 
calculated in section~\ref{MSFmethod} (open circles). 
The GGC metallicities listed in Table~\ref{GCgroups} have been transformed into global metallicities, 
[M/H], using the prescription of \citet{S93}: 

\begin{equation}
[M/H] = [Fe/H] + log(0.638f + 0.362)
\end{equation}

\noindent where $log f = [\alpha/Fe]$. For the $\alpha$-element
enhancement, based on previous literature estimates 
\citep{C96, SC96, Ve04, Kir08}, 
we assumed: $[\alpha/Fe] = +0.3$ $\pm$ 0.05 for clusters with $[Fe/H] <-1.0$, and $[\alpha/Fe] = +0.2$
$\pm$ 0.05 for clusters with $[Fe/H] \ge-1.0$. 
The effect that this $\pm$ 0.05 dex uncertainty has on the final relative ages will be discussed
at the end of this section.
The values of [M/H] in both ZW and CG scales are listed in 
columns 2 and 3 in Table~\ref{GCages}, respectively. We finally estimate the age of each cluster
based on the interpolated curves. 

While this procedure provides absolute ages for all of our clusters, it must be noted that absolute 
ages depend on the theoretical model as well as the adopted MSTO magnitude zeropoint. 
A much more detailed analysis is required to derive reliable absolute ages, and this will be done 
in a forthcoming paper. 
This paper concentrates only on relative ages, which are much less
dependent on the MSTO zero point.

In order to derive relative ages, we divided the target clusters 
into three groups: low$-$metallicity ([M/H]$<$-1.4),
intermediate$-$metallicity (-1.4$<$[M/H]$<$-0.8), and high$-$metallicity
([M/H]$>$-0.8).  Using the D07 models, we compute the mean age of the 
low$-$metallicity group as
$12.80\pm 0.17$ Gyr. The rms scatter of the 13 low$-$metallicity GGCs is equal to 0.6 Gyr.
The absolute age of each cluster was then divided by
this mean age, obtaining what we will call normalized age from here on.  
Columns 5 and 6 of Table~\ref{GCages} list the normalized ages
from the D07 models using the ZW and CG metallicity scales,
respectively.  

The adopted $\pm$ 0.05 dex uncertainty for the $\alpha$-element
enhancement translates into an uncertainty of $\sim$ 0.012 on the final
relative age. With the aim of taking into consideration this effect on the relative ages,
this quantity has been added quadratically to obtain the final relative age uncertainties 
listed in Table~\ref{GCages}.   

\subsection{Testing self$-$consistency}

In this section, the self$-$consistency of the MS$-$fitting procedure
and relative age determination is tested.  In particular, we are
interested in quantifying
the validity of matching up clusters with different metallicities on the
grounds of theoretical stellar evolution models. 

We created a set of 13 Gyr synthetic isochrones 
using the D07 models with metallicities 
similar to those of the selected reference clusters NGC 6304, NGC 6723,
NGC 6981, NGC 6681, NGC 6101 and NGC4590. That is, isochrones with
[M/H] = -0.5, -0.8, -1.0, -1.1, -1.5 and -1.8 have been considered.
We applied the same 
MS$-$fitting procedure to the isochrones as 
for the observational data (section~\ref{MSFmethod}).
The [M/H] = -1.1
isochrone, located in the center of the metallicity interval, was chosen for the MSTO zeropoint 
determination.  

\begin{figure}
\epsscale{1.00}
\plotone{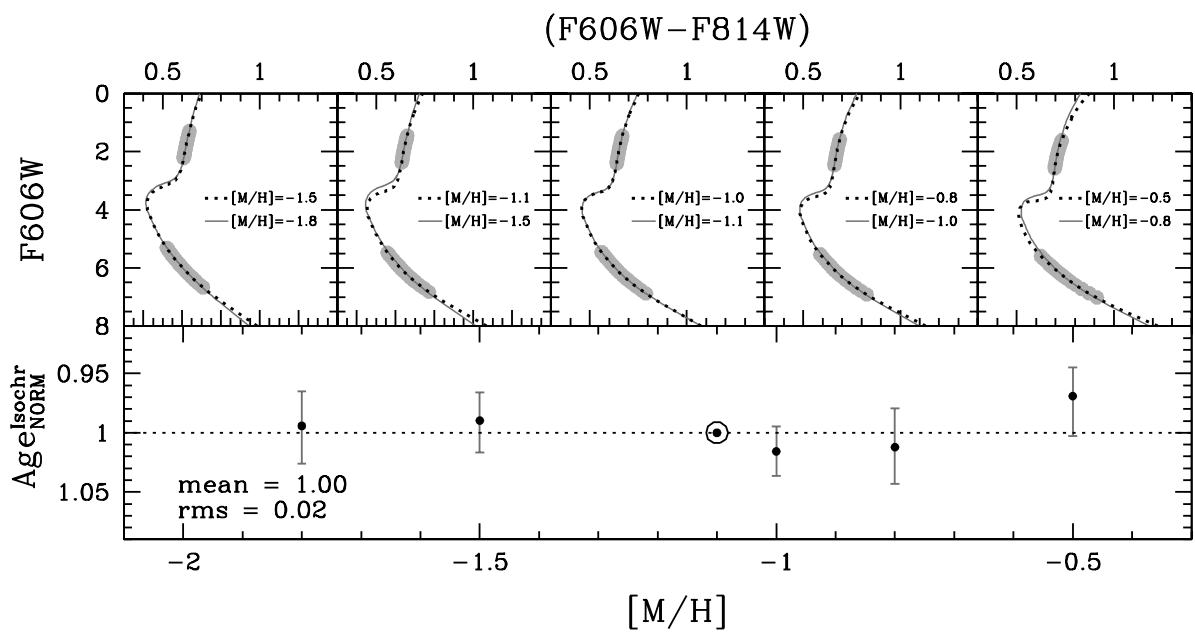}
\caption{Testing the used main sequence fitting procedure and relative ages determination 
using 13 Gyr synthetic isochrones. 
Upper panel shows the MS$-$fitting for 13 Gyr synthetic isochrones with the same metallicities as 
the adopted reference clusters ([M/H] = -0.5, -0.8, -1.0, -1.1, -1.5 and -1.8). Lower panel 
shows the obtained relative ages. The MSTO magnitude of the [M/H]=-1.1 
isochrone (encircled point) has been considered as MSTO zeropoint.
\label{isochrMSF}}
\end{figure}

Figure \ref{isochrMSF} 
summarizes the test results. Upper panel shows the MS$-$fitting for the 13 Gyr synthetic 
isochrones with the same metallicities as the adopted reference clusters. Lower 
panel shows the obtained relative ages results. The MSTO of the [M/H]=-1.1 
isochrone, that has been adopted as MSTO zeropoint, has been encircled.
Error bars represent the uncertainties derived from the MS$-$fitting, $\sigma_{\rm MSF}$.
It can be seen that the main sequence fitting procedure is self consistent, and that the
derived ages are similar to the input ages with a typical uncertainty of
less than 2\% over the entire metallicity range.  We thus conclude that our metallicity grouping should
not induce a significant bias in the relative age estimation.

\subsection{Comparison to other stellar evolution libraries}

\begin{figure}
\epsscale{1.00}
\plotone{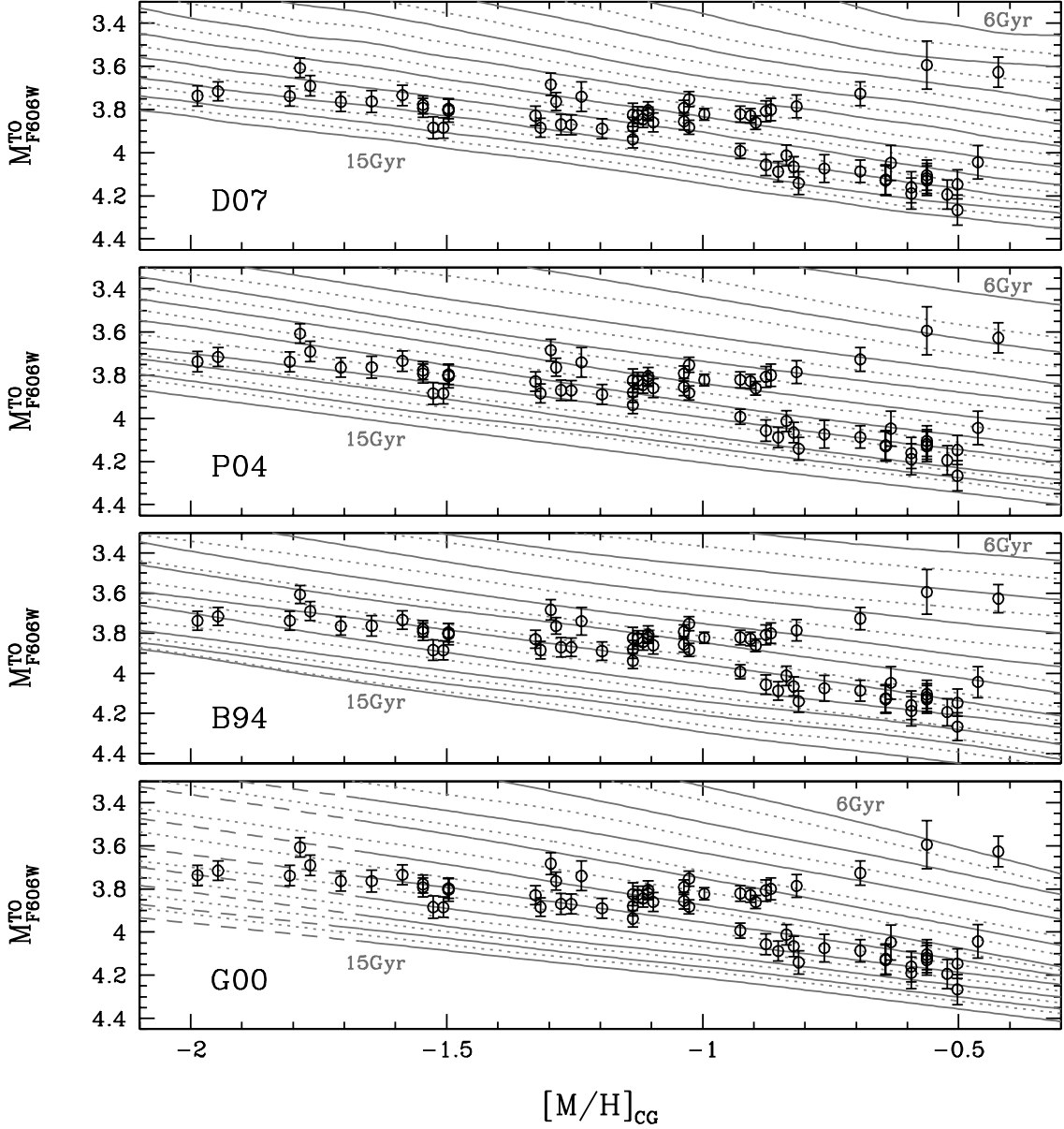}
\caption{MSTO magnitudes predicted by the D07, P04, B94, and G00 stellar 
evolution libraries. Lines represent $M^{\rm TO}_{\rm F606W}$ for different ages as a function 
of global metallicity, in 1 Gyr (solid lines) and 0.5 Gyr (dashed lines) age steps. 
The lower isochrone corresponds to an age of 15 Gyr and the upper one to 6 Gyr. 
In the case of G00 models, metallicities lower than Z=0.0004 have been extrapolated. 
Points represent observational data. \label{TOModels}}
\end{figure}

Figure~\ref{TOModels} shows the theoretical MSTO magnitudes derived using 
P04, B94, and G00 stellar evolution libraries. D07 model predictions have been also 
plotted for comparison. Lines represent $M^{\rm TO}_{\rm F606W}$ for different ages as a function 
of global metallicity, in 1 Gyr (solid lines) and 0.5 Gyr (dashed lines) age steps. Points 
represent our measured MSTO magnitudes as a function of metallicity in the CG 
scale. In the case of G00 models, metallicities lower than Z=0.0004 have been extrapolated. 

Using these theoretical grids, 
normalized ages have been derived following the procedure 
described in section \ref{relAges}. The corresponding average 
age of the low metallicity group, as derived from each isochrone set, 
has been used for normalization.

\begin{figure}
\epsscale{1.00}
\plotone{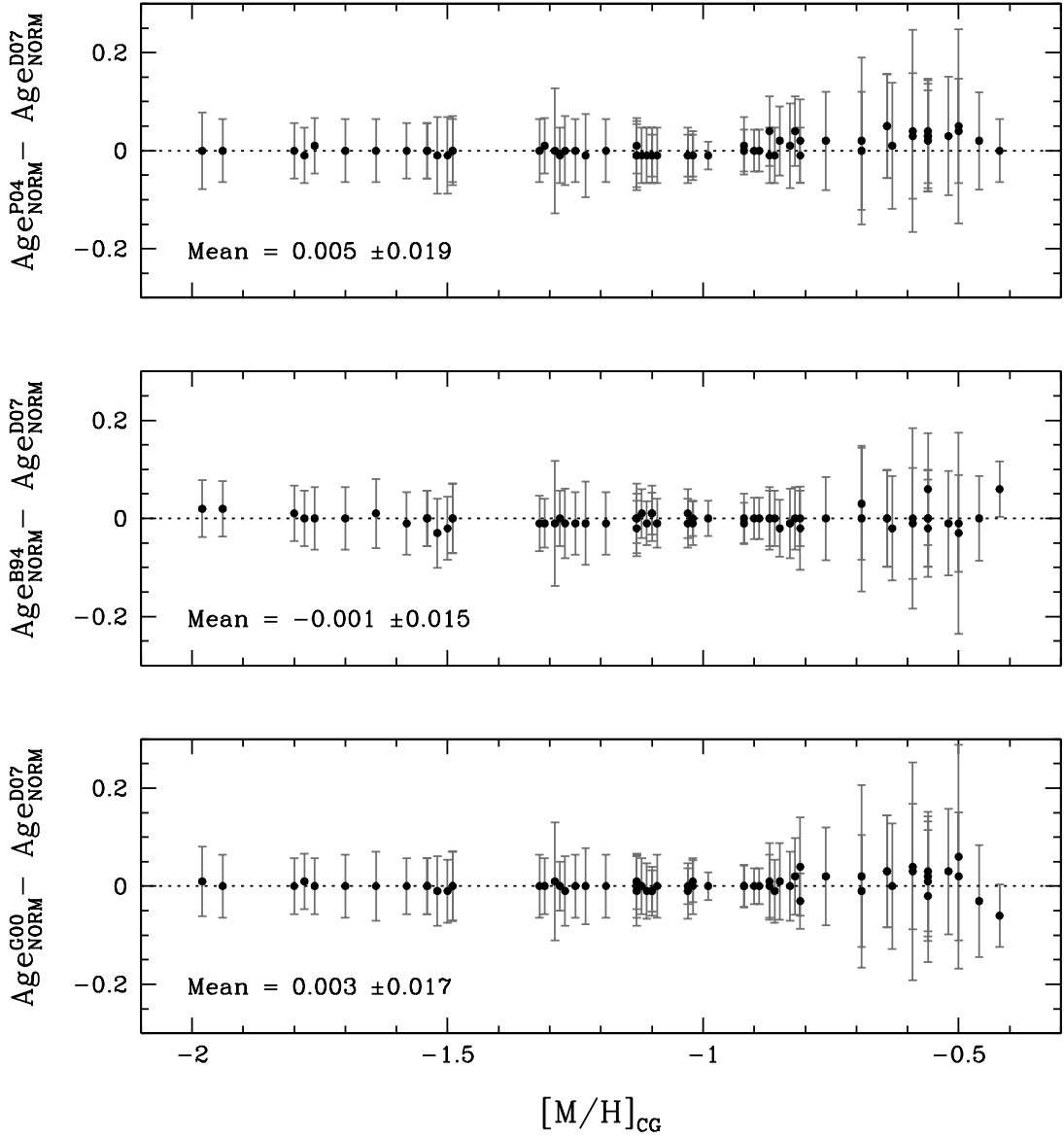}
\caption{Comparison of normalized ages derived from different 
sets of models in the CG metallicity
scale. The difference between normalized ages derived using P04, B94
and G00 models, respectively, and the D07 models are shown. It can be
seen that relative ages derived using the main sequence
fitting method are independent 
from the adopted model.  \label{modelComp}}
\end{figure}

Columns 7, 8 and 9 of Table~\ref{GCages} list the normalized ages derived from
the P04, B94 and G00 stellar evolution libraries, respectively, using the CG
metallicity scale. Figure~\ref{modelComp} shows the difference between the normalized ages
from the D07 models, and those from the P04, B94 and G00 models,
adopting the CG metallicity scale. The error bars are the 
quadrature of the relative age's uncertainties form the two 
corresponding models. Interestingly, the relatives ages 
are the same, within the uncertainties, independent
of the adopted theoretical library.

\subsection{Comparison to previous studies}

\begin{figure}
\epsscale{1.00}
\plotone{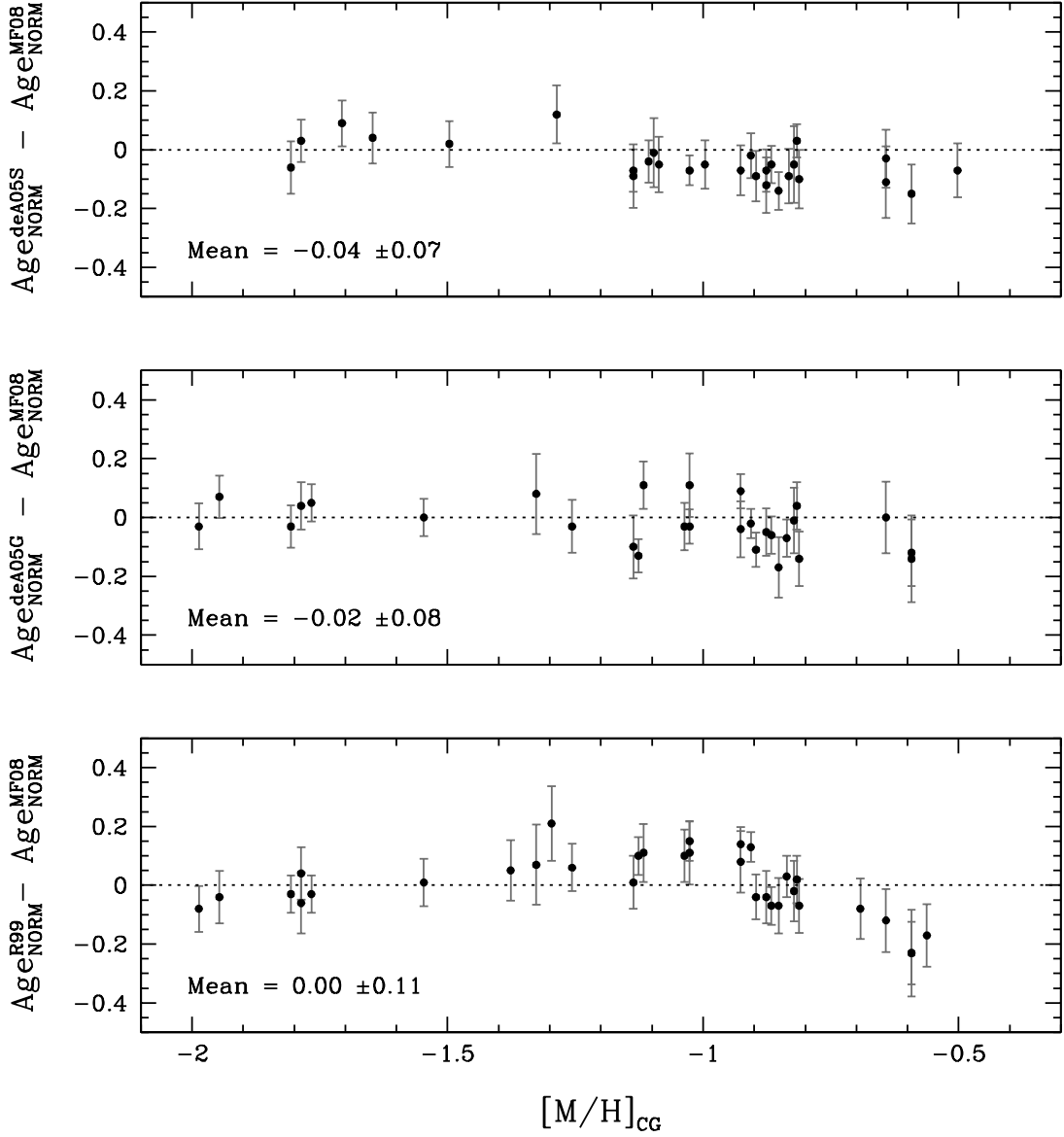}
\caption{Comparison with previous works using the CG metallicity
scale.  The differences between the normalized ages of
\citet{DeA05} (using 
HST snapshot -deA05S- and ground based -deA05G- data)
and \citet{R99} (R99) and the results derived 
in this paper by
using D07 models (MF08) are indicated. \label{perviousWorksComp}}
\end{figure}

Figure~\ref{perviousWorksComp} shows a comparison of our results to
previous work. The upper and middle panels show the difference between our results and the
normalized ages published by \citet{DeA05} using HST
snapshot -deA05S- and ground based -deA05G- data. The lower panel shows the difference between 
our results (MF08) and \citet{R99}'s normalized ages. The uncertainties are the quadrature of our listed uncertainties 
and those published in the \citet{DeA05} and \citet{R99} studies.
The \citet{DeA05} ages are 
consistent with ours within the error bars and no appreciable age trend is observed. 
On the other hand, some trend is seen when comparing our results to those 
by \citet{R99}: \citet{R99} find marginally older ages for cluster with $-1.4<$[M/H]$<-0.9$ and 
younger ages for high metallicity clusters. This discrepancy is in part due to the 
different methods applied to measure the observed cluster parameters \citep[see 
discussion in][and in particular their Figure~7]{DeA05}, and in part to the different 
evolutionary models adopted for the age determination \citep [][uses the P04 
models, which we have shown to give consistent results with our adopted D07 
models]{DeA05} .

\subsection{The special case of NGC 0288 and NGC 0362}\label{n228n362}

In this study MSTO's magnitude is measured for each GGCs in our database, and then obtained
magnitudes are transformed into ages using a set of theoretical stellar evolution models.
During this transformation, canonical chemical abundances are assumed for all GGCs. We
are aware of the fact that if a particular cluster has different chemical abundances, for example, 
an anomalous CNO content, then its age determination
could be affected by this effect. It is worth mentioning that age uncertainties could be underestimated 
in this study because of the uncertainties coming from the chemical inhomogeneities
which, to date, are impossible to quantify.

In particular, our results indicate that NGC 0288 and NGC 0362 have the same age
within $\pm$ 0.9 Gyr, 
while previous works \citep{SD90,B01} have found NGC 0288 to be 1-2 Gyr older 
than NGC 0362. 
The relative ages from these studies come from a comparison of all the evolutionary sequences (MS, 
SGB, HB, etc.), and not on a measurement of the absolute magnitude of the 
MSTO, as in the present paper. Therefore, we are planning a forthcoming paper in 
which we will, among other things, consider the case of these two clusters in more detail.
In this context, it is worth mentioning that the present study does not take into consideration the distribution of stars
along the HB, or finer features such as the RGB tip luminosity, the RGB bump, 
AGB bump and multiple populations. 

A similar argument could be used for the case of NGC 5272 and NGC 6205, for 
which we obtain identical age, while previous works \citep{VBS90, Ch96} have found NGC 6205 
to be 1-2 Gyr older than NGC 5272.

\section{Discussion}\label{discussion}

\begin{figure}
\epsscale{1.00}
\plotone{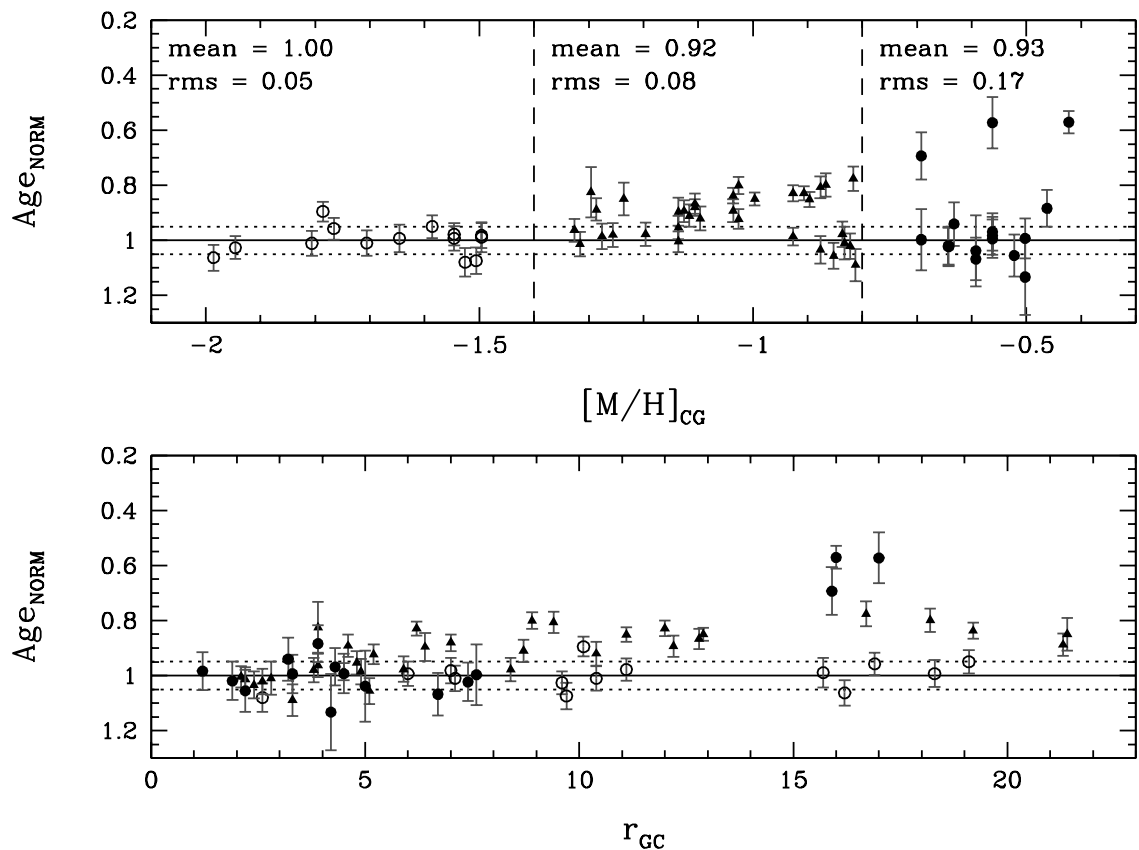}
\caption{Globular cluster normalized ages as a function of [M/H] in
the CG metallicity scale (upper pannel), and as a function of
galactocentric distance ($r_{\rm GC}$ kpc, lower pannel). These results
have been derived using the D07 stellar evolution library. Open circles,
filled triangles and filled circles represent GCs within the low$-$,
intermediate$-$ and high$-$metallicity groups, respectively. For each
of the three metallicity groups, mean age and rms
is indicated.  See text for
details. \label{ChaboyerMSFmethod}}
\end{figure}

\begin{figure}
\epsscale{1.00}
\plotone{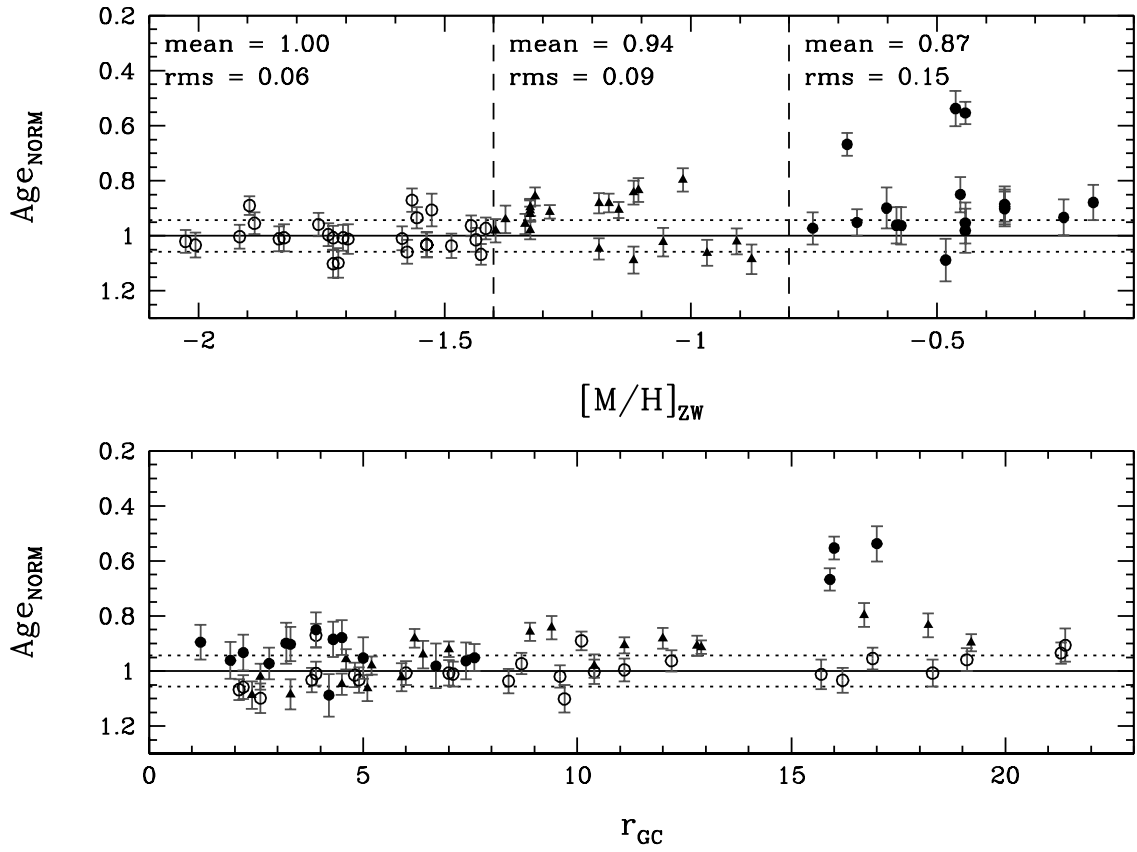}
\caption{Same as Figure \ref{ChaboyerMSFmethod}, but for the ZW metallicity 
scale. \label{ChaboyerMSFmethodZW}}
\end{figure}

Figure~\ref{ChaboyerMSFmethod} shows the GGC normalized ages derived using the
D07 stellar evolution library as a function of [M/H] in the CG metallicity scale (upper panel), 
and as a function of the galactocentric distance ($r_{\rm GC}$, lower panel). The low$-$, 
intermediate$-$ and high$-$metallicity subgroups are 
separated by long dashed lines, and
clusters belonging to the different metallicity groups are plotted with open 
circles (low$-$), filled triangles (intermediate$-$) and filled circles (high$-$metallicity). 
The mean and dispersion of the low$-$metallicity clusters' age are marked with solid and 
dashed lines, respectively. For each of the three metallicity groups,
mean age and rms is indicated. Figure \ref{ChaboyerMSFmethodZW} 
shows the same data as Figure \ref{ChaboyerMSFmethod}, but for the ZW metallicity scale. 

Overall, we notice an increase in age dispersion (which is model
independent) with the metallicity. However, we can look at the results
shown in the upper panels of Figures~\ref{ChaboyerMSFmethod} and
\ref{ChaboyerMSFmethodZW} in another way. They show two branches in
the age$-$metallicity relation of GGCs. On one hand, a 'young' branch
showing a clear age$-$metallicity relation, age decreasing at higher
metallicities. On the other hand, a largely coeval 'old' branch. The
'old' branch shows an apparent relative age dispersion of $\sim$0.05 and no
age$-$metallicity relation.  It is worth mentioning that while the
age$-$metallicity relation may be model dependent (though it is
somehow reassuring that the four most recent theoretical libraries
provide consistent results on this respect) the age
dispersion$-$metallicity relation is not model dependent.  From now
on, we will divide the clusters in two groups that we call 'young' and
'old'. GGCs with total metallicity higher than -1.4 and normalized age
younger than 0.95, that is one sigma smaller than the mean of the
low$-$metallicity clusters' age, are considered members of the 'young'
group.  The remaining GGCs are considered members of the 'old' group.
The only exception is NGC 6441, a very high metallicity cluster which
is much closer to the old branch than the young one, so it is included
in the old group.  Column 10 in Table~\ref{GCages} lists whether the
target cluster belongs to the old or young group.

\begin{figure}
\epsscale{0.9}
\plotone{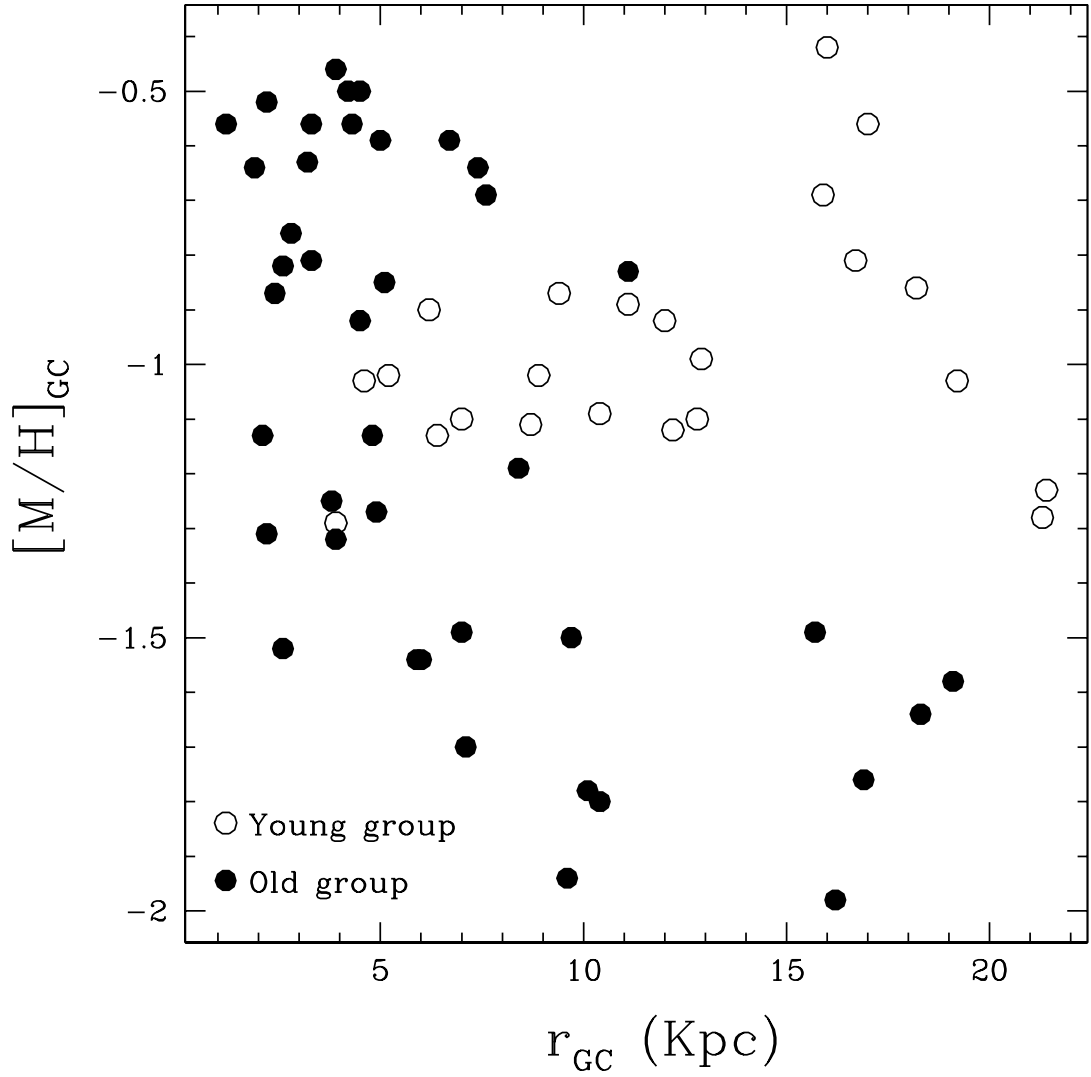}
\caption{Globular cluster's [M/H] in the CG metallicity scale as a function of the galactocentric distance. Open and 
filled circles represent clusters in the young and old groups, respectively.  
\label{MH_rGC}}
\end{figure}

An age trend with galactocentric distance is seen in the lower panels of
Figures~\ref{ChaboyerMSFmethod}  and \ref{ChaboyerMSFmethodZW}. 
The fraction of young group's clusters increases significantly as the galactocentric 
distance increases. As a result, age dispersion increases at increasing galactocentric distance.
Other authors have commented on the increased dispersion in ages at
larger galactocentric distances.  \citep[e. g. ][which is one of the first to state the empirical problem clearly]{R96}.
Also in these panels, a clear metallicity trend with galactocentric distance can be observed 
for the old group of clusters, with an increasing metallicity at lower galactocentric distances.
This trend is explicitly shown in Figure~\ref{MH_rGC}, which plots GGC's [M/H] in the CG metallicity scale as
a function of the galactocentric distance. Open and filled circles represent clusters in the young 
and old groups, respectively. 

In the following, the age dispersion of the old clusters group, the age$-$metallicity relation of the young clusters group
and the age trend with galactocentric distance are discussed in detail.

\subsection{The age dispersion of the old clusters group}

Figures~\ref{ChaboyerMSFmethod} and \ref{ChaboyerMSFmethodZW} indicate
that the mean normalized age of the clusters in the old group is the same for all metallicities.
The relative age dispersion of this group is $\sim$0.05. If an absolute age
of 12.8 Gyr (from the D07 models) is assumed as the mean age of these clusters,
age dispersion would be $\sim$0.6 Gyr. We note that if we assume that all
sources of uncertainty have been taken into account in the error bars, then 
at least part of this dispersion must be real. The average normalized age uncertainty among the 
clusters in the old group is 0.04, which is approximately 0.5 Gyr.  Subtracting this quadratically from 
the observed dispersion, we find an intrinsic dispersion of 0.03 in relative age for the clusters in the 
old group. Using 12.8 Gyr for the age of this group, the intrinsic dispersion is $\sim$0.4 Gyr. 

These results are consistent with a scenario in which the old group of clusters formed within a 
fast assembling process of the halo, lasting $\sim$0.8 Gyr and which should also account for an
increasing metallicity towards the center of the Galaxy. This implies that 
the chemical enrichment of the protogalactic cloud was faster than the 
assembling timescale. 
Regarding the nature of such assembling process, it is interesting to note the following. \citet{W99} estimate the 
mass and scalelength of the Milky Way dark matter halo to be $\sim1.9\times10^{12}$ M$_\odot$ 
and 170 kpc, respectively. 
The free-fall time of a homogeneous sphere of those mass and radius is $\sim0.84$ Gyr, 
similar to the age range we find here for the old clusters. In other words, the age dispersion 
of old globular clusters is not in contradiction with the formation from the colapse of a 
single protosystem, resembling the model proposed by \citet{ELS62}.  
On another side, the standard $\Lambda$CDM scenario would 
predict a general star formation, including star clusters, in protogalaxy building blocks 
of mass $\sim10^8$ M$_\odot$ before reionization \citep{Moore06}. These blocks 
merge together afterwards to form larger galaxies, and would, in principle, be compatible 
with the old, coeval cluster population that we observe. More difficult seems to account for the metallicity gradient (Fig.~\ref{MH_rGC}) under this scenario. A more detailed analysis is 
necessary, but it is important to note that any successful galaxy formation scenario 
must account for the existence of a large nubmer of old, coeval globular clusters in the 
Milky Way.

\subsection{The Age$-$metallicity relation of the young clusters group}

Clusters in the young group, however, show a clear age$-$metallicity relationship, with younger clusters 
being more metal rich than older ones. The nature of this ``young branch" 
can perhaps be better understood by looking at the GGCs associated
with known or putative accreted dwarf galaxies: Sagittarius, Canis Major and
Monoceros.

\citet{I95} discovered the accreted Sagittarius dwarf
galaxy and its GC system (Terzan 7, Terzan 8, Arp 2, and M54).
\citet{Din00} and \citet{B03} have since then argued, respectively, that Pal 12 and
NGC 4147 are part of the extended tidal stream.

The Monoceros Ring was discovered by \citet{N02}, and
it has been proposed to be the tidal stream of the accreted Monoceros
galaxy. \citet{C03} and \citet{F04} identified five possible GC
candidates of this accreted galaxy: NGC 2298, NGC 2808, NGC 5286, Pal
1, and BH 176.  In addition, \citet{M04} identify four GGCs -- NGC 1851, 
NGC 1904, NGC 2298 and NGC 2808 -- possibly associated with
a stellar overdensity in Canis Major, which they suspect to be a dwarf galaxy.
However, the nature of these two structures is controversial.  The existence of the Canis 
Major dwarf galaxy has been seriously questioned by \citet{Mom04, Mom06}. 
In \citet{Mom06}, the Canis Major stellar over$-$density has been 
completely accounted for as the effect of the Milky Way disk warp (an ubiquitous 
property of all massive galaxies). In addition, \citet{Mom06} have shown that also the Monoceros Ring 
seems to reflect the signature of another Galactic disk property; the flaring (the increase in 
scale-height as a function  of Galactocentric distance) of the outer disk. 

\begin{figure}
\epsscale{0.9}
\plotone{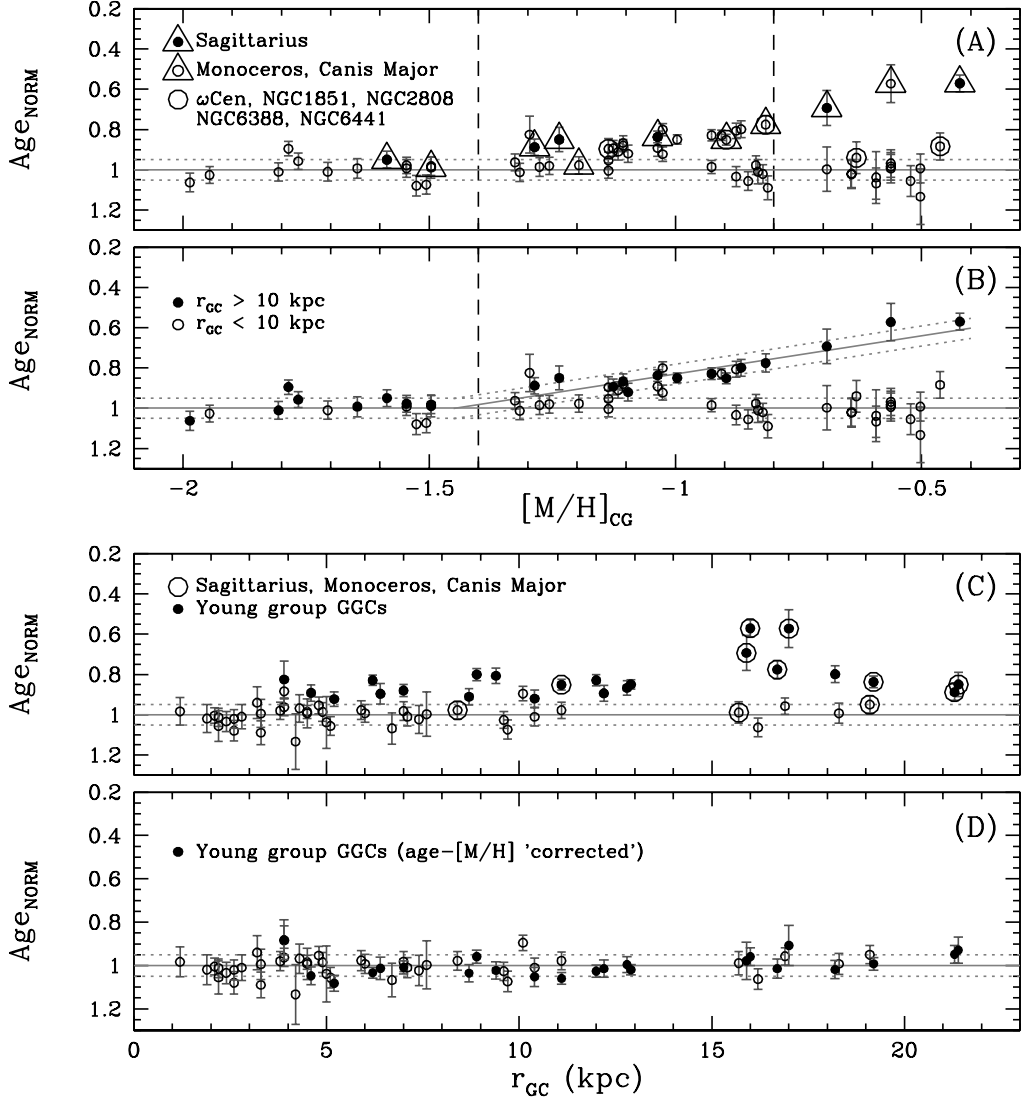}
\caption{Age$-$metallicity relation using the CG metallicity scale (A and B) and
normalized ages versus galactocentric distance (C and D). In A, clusters 
associated with Sagittarius, Monoceros and Canis Major have been marked, together with
multiple stellar population clusters. In B, GGCs 
with $r_{\rm GC} < 10 $ kpc are represented with open circles, while those 
with $r_{\rm GC} > 10 $ kpc are plotted with filled circles. 
A least squares fit to the young group's age$-$metallicity relation is also plotted. In C, Sagittarius, Monoceros and
Canis Major' clusters are also marked. Finally, D shows the same as C, but here ages of clusters
in the young group have been age$-$metallicity 'corrected' using the age$-$metallicity relation shown in B. 
See text for details. \label{ageMetallicity}}
\end{figure}

Figures~\ref{ageMetallicity}A and \ref{ageMetallicity}B show the age$-$metallicity relation, 
using the CG metallicity scale\footnote{Because the results are equivalent to those using 
the ZW metallicity scale, we only consider the CG scale after this point.}. 
Sagittarius, Monoceros and Canis Major's clusters have been marked. 
It is interesting that, for metallicities higher than [M/H] = -1.4, all of Sagittarius, Monoceros 
and Canis Major's accreted clusters fall in the young group, with the exception of NGC 5286, 
and they follow the same age$-$metallicity relation as the rest of the young group's clusters. 
This result suggests a different origin for the old and young groups of clusters.
It is worth mentioning that there is a number of clusters, other than the Sagittarius, Monoceros 
and Canis Major's clusters among the young sample.

Another interesting question 
is related to the recent discovery that GGCs are not simple single 
stellar populations. The high mass and multiple 
stellar populations of $\omega$Cen has led to speculation that
it is the remnant nucleus of an accreted Milky Way satellite galaxy
\citep{L99, V07}. But $\omega$Cen is not a unique case. 
NGC 2808 \citep{P07}, NGC 1851 \citep{M07}, NGC 6388 \citep{P08} and
NGC 6441 \citep{CD07} are four massive clusters hosting multiple stellar populations. 
\citet{F93} suggested that GGCs form as the nuclei of dwarf galaxies in the early universe, 
and are accreted as their host galaxies merge onto larger structures. In this context,
these peculiar GGCs would be good candidates to be the
remnants of accreted satellites. Multiple population clusters have been also 
marked in Figure~\ref{ageMetallicity}A. 

We note that multiple stellar population clusters have broadened MSTOs 
and at least a fraction of the stars in these clusters have
rather anomalous chemical compositions. For this reason, we should be hesitant
in adopting their relative ages obtained in this study.  A
much more detailed analysis is needed for these peculiar clusters.  
\citet{V07} and \citet{C08} present particularly illuminating discussions of 
the difficulties in estimating ages for the multiple populations in $\omega$Cen and
NGC 1851, respectively. 

Figure~\ref{ageMetallicity}B shows the same Figure~\ref{ageMetallicity}A,
but here GGCs with $r_{\rm GC} > 10$ kpc have are represented with filled circles, 
while those with $r_{\rm GC} < 10 $ kpc are plotted with open circles. Interestingly, all 
intermediate$-$ and high$-$metallicity clusters with $r_{\rm GC} > 10$ kpc
fall on the young group of clusters. 

In order to characterize the young group's age$-$metallicity relation, 
a least squares fit has been performed, and results are 
also shown in Figure~\ref{ageMetallicity}B.
The age$-$metallicity relation observed in the young group of clusters can be described
by the equation:

\begin{equation}\label{AgeZrelation}
{\rm Age}_{\rm NORM} = -0.38 [{\rm M/H}] + 0.45
\end{equation}
 
\noindent This relation is plotted with a solid line in Figure~\ref{ageMetallicity}B.
The dispersion of the clusters' relative ages with respect to the previous equation is also 0.05, and it is 
also represented in Figure~\ref{ageMetallicity}B (dashed lines).  
Following a similar argument as for the old group, if an absolute age of 12.8 Gyr is 
assumed as the mean of the clusters in the old group, 
the young group's age dispersion with respect to the previous age$-$metallicity relation 
would be $\sim$0.6 Gyr. The average normalized age uncertainty among the 
clusters in the young group is 0.04, which is approximately 0.5 Gyr.  Subtracting this
quadratically from the observed dispersion, 
we find an intrinsic dispersion with respect to its age$-$metallicity relation of 0.03 in 
relative age, or $\sim$0.4 Gyr.

In summary, our results are consistent with the clusters of the young group having been formed in a different process spanning an interval of time as long as 0.45 in relative age, or $\sim$6 Gyr,
and resulting in a group of GCs with a clear age$-$metallicity relation. 
It is very tempting to argue that the origin of this second group of clusters is related to 
their formation within Milky Way satellite galaxies that were later accreted. 
However, the reason why the age$-$metallicity relation of the GGCs associated 
with the Sagittarius dwarf galaxy, as well as Monoceros and Canis Major, is similar to 
that of the rest of 'accreted' clusters should be investigated. 
Perhaps all of the young group clusters share the same origin, which would require 
an association between Sagittarius, Monoceros and Canis Major, or alternatively, 
all dwarf galaxy systems may share a common (or very similar) star formation and 
metal enrichment history, and hence a common age$-$metallicity relation, which seems unlikely.

\subsection{Age trend with galactocentric distance}

With the aim of analyzing in more detail the observed age trend with galactocentric 
distance, GGCs in the old and young groups are considered separately. 
Figure~\ref{ageMetallicity}C shows normalized ages versus galactocentric distance
(same as Figure~\ref{ChaboyerMSFmethod}, lower panel), but now clusters in the old group 
are represented with open circles and clusters in the young group with
filled circles. 
Besides, Sagittarius, Monoceros and Canis Major's clusters are shown as encircled points.

It can be seen that if the old group is considered, an age trend with galactocentric distance is not 
observed. The age's variance remains constant with the galactocentric distance for this group.

\begin{figure}
\epsscale{1.00}
\plotone{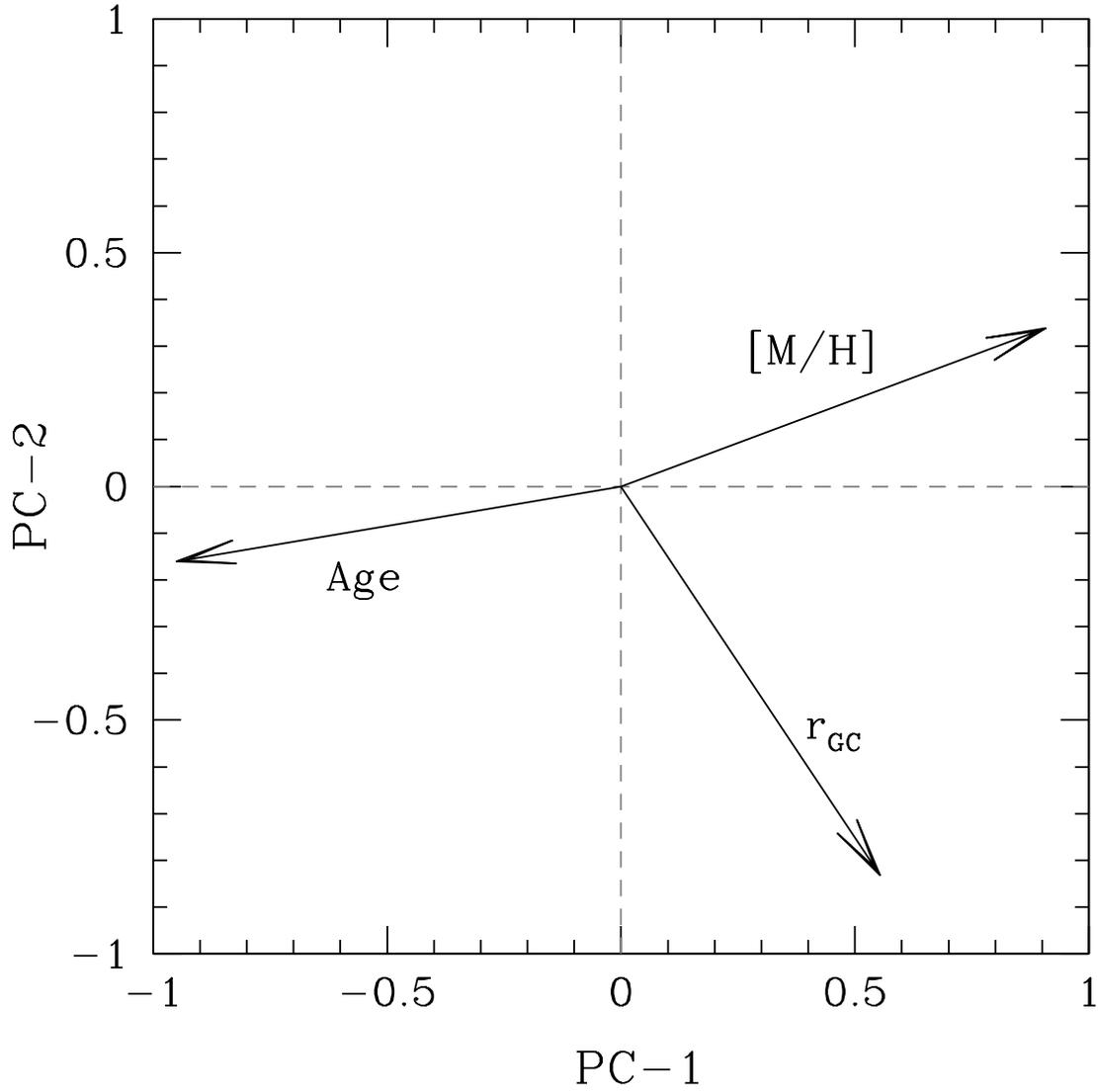}
\caption{Young group principal$-$component (PC) analysis: Relations between original data and principal components on the first two 
principal components plane. It can be seen that the 
first PC is strongly correlated with [M/H] and age, but weakly with $r_{GC}$, while the second 
PC carries on most of the $r_{GC}$ variance. See text for details. \label{pcaResults}}
\end{figure}

On the other hand if the young group is taken into account, it can be seen
that the age's variance increases with galactocentric distance. In order to
determine if this variance increase depends either on the galactocentric 
distance or the metallicity, a three dimensional principal$-$component (PC) analysis 
has been done considering the clusters in the young group only. The 
obtained eigenvalues are 2.03, 0.83 and 0.14. The first PC, of eigenvalue 2.03, accounts for 
68$\%$ of the total variance. According to the commonly used average or eigenvalue-one criterion,
only this PC should be retained. Figure~\ref{pcaResults} shows the relations between the original 
data and PCs projected on the two first PCs plane. On the one 
hand it can be seen that the 
first PC is strongly correlated with [M/H] and age, but weakly with $r_{GC}$, while the second 
PC carries on most of the $r_{GC}$ variance. On another 
hand it is clear that young cluster 
age dispersion can be explained by an age$-$metallicity relation and that 
age is not related to $r_{GC}$. 

In fact, the effect of metallicity on age of clusters 
in the young group can be eliminated by subtracting the fitted age$-$metallicity relation 
(equation~\ref{AgeZrelation}) to the cluster's ages. By doing this, age$-$metallicity 'corrected' ages
are obtained. Figure~\ref{ageMetallicity}D shows the same as Figure~\ref{ageMetallicity}C, but here
age$-$metallicity 'corrected' clusters are represented with solid circles. It can be clearly seen
that most of the age dispersion observed in Figure~\ref{ageMetallicity}C disappears, and that
age is not related to $r_{GC}$

In summary, it can be concluded that a significant part of the age
dispersion present in Figure~\ref{ageMetallicity}C for $r_{GC} \le 15$ kpc and 
most of that for $r_{GC} > 15$ can be explained
by a strong age$-$metallicity relation for the clusters of the young group. 

It is worth mentioning that our database is biased in distance. No outer$-$halo clusters, 
with galactocentric distances larger than $\sim$20 kpc, have been considered for this study. 
So the analysis of the galactic halo's age structure is limited by the galactocentric distance range of the 
database. Given that outer$-$halo clusters (Pal 3, Pal 4, Pal 14, Eridanus, AM-1) have 
younger ages \citep{Sar97, Stet99, DSY08}, it is possible that our age range might be a lower limit
on the actual range.

\section{Conclusions}\label{conclusions}

Normalized ages have been derived for a sample of 64 GGCs using the stellar
evolution models of D07 and both ZW and CG metallicity scales.  We have also performed
an analysis using the stellar models of 
P04, B94 and G00.  The result is the most extensive and precise database of normalized ages
so far produced . Our results are:

\begin{itemize}

\item We find that we are able to measure relative ages to a formal precision of 2-7\%.
The relative ages are independent of the choice of theoretical model.  Four independent sets
of stellar evolution libraries (D07, P04, B94, G00) produce essentially identical results.

\item We find that the GGCs fall into two well-defined groups.  The first one represents a 
population of old clusters that have the same age, a dispersion of 
5\% in relative age and no age$-$metallicity relation. If we assume an
absolute age of 12.8 Gyr, the absolute age dispersion of the old clusters is $\sim$0.6 Gyr. 
Accounting for the measurement uncertainties, 
we obtain an intrinsic age dispersion of $\sim$0.4 Gyr. The second group of clusters shows
a clear age$-$metallicity relation, with young clusters being more
metal rich than older ones. 
Also in this case, the intrinsic age dispersion with respect to its age$-$metallicity relation 
is $\sim$0.4 Gyr. 
Besides, there is no age dispersion$-$metallicity relation if we look at the two 
samples (old and young) separately.

\item These results are consistent with a scenario in which the formation 
of the Milky Way CCG system took place in two phases. The first one produced 
clusters with ages within a range of $\sim$0.8 Gyr. This age range is compatible with the 
timescale for the collapse of a protogalaxy of the same mass and scalelength as 
that of the Milky Way dark matter halo. In other words, the age dispersion of the old group of globular 
clusters is not in contradiction with the formation from the colapse 
of a single protosystem, resembling the model proposed by \citet{ELS62}. 
The standard $\Lambda$CDM scenario would also predict a 
significant star formation in protogalactic building blocks before reionization, 
which, after merging, would produce an old, coeval population of globular clusters 
in a large galaxy like the Milky Way. However, to account for the galactocentric metallicity gradient of this group seems difficult in the context of this mechanism.
The second phase spanned a time interval as long as $\sim$6 Gyr
and resulted in a group of GGCs with a clear age$-$metallicity 
relation. It is very tempting to argue that the origin of this second group of 
clusters is related to their formation within Milky Way satellite galaxies that 
were later accreted. However, this would not account for the fact that the clusters of this group share the same age$-$metallicity relation.

\item  We find that the increasing dispersion in age with galactocentric distance 
is explained by the age$-$metallicity relation of GGCs. 

\item These results are independent of the assumed metallicity scale.

\item It is worth mentioning that our database is biased by distance. No outer 
halo clusters, with galactocentric 
distances larger than $\sim$20 kpc, have been considered. The analysis 
of the galactic halo's age structure is therefore limited by the galactocentric distance range of the 
database.

\end{itemize}

\begin{acknowledgements}
Support for this work (proposal GO-10775) was provided by NASA through a grant from the 
Space Telescope Science Institute, which is operated by the Association of Universities for 
Research in Astronomy, Inc., under NASA contract NAS5-26555. AMF has been 
partially supported by the Gran Telescopio Canarias Postdoctoral Fellowhip through the 
University of Florida and the Education and Research Ministry of Spain's 'Juan de la Cierva' 
postdoctoral position. This work has been financially supported by the Instituto de 
Astrof\' \i sica de Canarias (grant P3-94) and the Education and Research Ministry of Spain 
(grant PNAYA2004-06343). GP and APM acknowledge partial support by the Agenzia 
Spaziale Italiana.

\end{acknowledgements}

\clearpage

\clearpage

\clearpage

\clearpage

\clearpage

\clearpage

\clearpage

\clearpage

\clearpage

\clearpage

\clearpage

\clearpage

\clearpage

\clearpage

\clearpage

\begin{deluxetable}{lclccc}
\tabletypesize{\scriptsize}
%\rotate
\tablecaption{Globular cluster parameters \label{GCgroups}}
\tablewidth{0pt}
\tablehead{
\colhead{Ref. cluster} & \colhead{Metallicity group} & 
\colhead{Name} & \colhead{[Fe/H]$_{\rm ZW}$} & \colhead{[Fe/H]$_{\rm CG}$}  & \colhead{r$_{\rm GC}$}}
\startdata
NGC 6304  &   -0.8$\le$[Fe/H]$_{\rm CG}<$-0.3  &  NGC 0104      &     -0.71  &  -0.78  &  7.4  \\
          &                                      &  NGC 5927      &     -0.32  &  -0.64  &  4.5  \\
          &                                      &  NGC 6304      &     -0.38  &  -0.66  &  2.2  \\
          &                                      &  NGC 6352      &     -0.50  &  -0.70  &  3.3  \\
          &                                      &  NGC 6366      &     -0.58  &  -0.73  &  5.0  \\
          &                                      &  NGC 6388      &     -0.74  &  -0.77  &  3.2  \\
          &                                      &  NGC 6441      &     -0.59  &  -0.60  &  3.9  \\
          &                                      &  NGC 6496      &     -0.50  &  -0.70  &  4.3  \\
          &                                      &  NGC 6624      &     -0.50  &  -0.70  &  1.2  \\
          &                                      &  NGC 6637      &     -0.72  &  -0.78  &  1.9  \\
          &                                      &  NGC 6838      &     -0.58  &  -0.73  &  6.7  \\
          &                                      &  LYNG\AA7      &     -0.62  &  -0.64  &  4.2  \\
          &                                      &  PAL 1         &     -0.60  &  -0.70  & 17.0  \\
          &                                      &  TERZAN7       &     -0.05  &  -0.56  & 16.0  \\
\hline
NGC 6723  &   -1.1$\le$[Fe/H]$_{\rm CG}<$-0.8  &  NGC 0362      &     -1.33  &  -1.09  &  9.4  \\
          &                                      &  NGC 1261      &     -1.32  &  -1.08  & 18.2  \\
          &                                      &  NGC 1851      &     -1.23  &  -1.03  & 16.7  \\
          &                                      &  NGC 6121      &     -1.27  &  -1.05  &  5.9  \\
          &                                      &  NGC 6171      &     -1.09  &  -0.95  &  3.3  \\
          &                                      &  NGC 6362      &     -1.18  &  -0.99  &  5.1  \\
          &                                      &  NGC 6652      &     -0.99  &  -0.97  &  2.8  \\
          &                                      &  NGC 6717      &     -1.33  &  -1.09  &  2.4  \\
          &                                      &  NGC 6723      &     -1.12  &  -0.96  &  2.6  \\
          &                                      &  E3            &     -0.80  &  -0.83  &  7.6  \\
          &                                      &  PAL 12        &     -0.82  &  -0.83  &  15.0  \\
\hline                                      
NGC 6981  &   -1.3$\le$[Fe/H]$_{\rm CG}<$-1.1  &  NGC 0288      &     -1.40  &  -1.14  & 12.0  \\
          &                                      &  NGC 2808      &     -1.36  &  -1.11  & 11.1  \\
          &                                      &  NGC 3201      &     -1.53  &  -1.24  &  8.9  \\
          &                                      &  NGC 5904      &     -1.38  &  -1.12  &  6.2  \\
          &                                      &  NGC 6218      &     -1.40  &  -1.14  &  4.5  \\
          &                                      &  NGC 6254      &     -1.55  &  -1.25  &  4.6  \\
          &                                      &  NGC 6584      &     -1.51  &  -1.30  &  7.0  \\
          &                                      &  NGC 6715      &     -1.54  &  -1.25  & 19.2  \\
          &                                      &  NGC 6752      &     -1.54  &  -1.24  &  5.2  \\
          &                                      &  NGC 6981      &     -1.50  &  -1.21  & 12.9  \\
\hline
NGC 6681  &   -1.5$\le$[Fe/H]$_{\rm CG}<$-1.3  &  NGC 4147      &     -1.77  &  -1.50  & 21.3  \\
          &                                      &  NGC 5139      &     -1.59  &  -1.35  &  6.4  \\
          &                                      &  NGC 5272      &     -1.66  &  -1.34  & 12.2  \\
          &                                      &  NGC 5286      &     -1.70  &  -1.41  &  8.4  \\
          &                                      &  NGC 5986      &     -1.65  &  -1.35  &  4.8  \\
          &                                      &  NGC 6093      &     -1.75  &  -1.47  &  3.8  \\
          &                                      &  NGC 6205      &     -1.63  &  -1.33  &  8.7  \\
          &                                      &  NGC 6535      &     -1.78  &  -1.51  &  3.9  \\
          &                                      &  NGC 6656      &     -1.75  &  -1.49  &  4.9  \\
          &                                      &  NGC 6681      &     -1.64  &  -1.35  &  2.1  \\
          &                                      &  NGC 6934      &     -1.54  &  -1.32  & 12.8  \\
          &                                      &  NGC 7089      &     -1.61  &  -1.31  & 10.4  \\
          &                                      &  ARP 2         &     -1.74  &  -1.45  & 21.4  \\
\hline
NGC 6101  &   -1.8$\le$[Fe/H]$_{\rm CG}<$-1.5  &  NGC 2298      &     -1.91  &  -1.71  & 15.7  \\
          &                                      &  NGC 4833      &     -1.92  &  -1.71  &  7.0  \\
          &                                      &  NGC 6101      &     -1.95  &  -1.76  & 11.1  \\
          &                                      &  NGC 6144      &     -1.81  &  -1.56  &  2.6  \\
          &                                      &  NGC 6397      &     -1.94  &  -1.76  &  6.0  \\
          &                                      &  NGC 6541      &     -1.79  &  -1.53  &  2.2  \\
          &                                      &  NGC 6809      &     -1.80  &  -1.54  &  3.9  \\
          &                                      &  TERZAN8       &     -1.97  &  -1.80  & 19.1  \\
\hline
NGC 4590  &   -2.3$\le$[Fe/H]$_{\rm CG}<$-1.8  &  NGC 4590      &     -2.11  &  -2.00  & 10.1  \\
          &                                      &  NGC 5024      &     -2.04  &  -1.86  & 18.3  \\
          &                                      &  NGC 5053      &     -2.10  &  -1.98  & 16.9  \\
          &                                      &  NGC 5466      &     -2.22  &  -2.20  & 16.2  \\
          &                                      &  NGC 6341      &     -2.24  &  -2.16  &  9.6  \\
          &                                      &  NGC 6779      &     -2.20  &  -2.00  &  9.7  \\
          &                                      &  NGC 7078      &     -2.13  &  -2.02  & 10.4  \\
          &                                      &  NGC 7099      &     -2.05  &  -1.92  &  7.1  
          
\enddata
\end{deluxetable}

\clearpage

\begin{deluxetable}{ccc}
\tabletypesize{\scriptsize}
%\rotate
\tablecaption{External cross-check for the $\sigma^{2}_{\rm MRL}$ determination \label{sigmaMRLtest}}
\tablewidth{0pt}
\tablehead{\colhead{Name} & \colhead{$\sigma_{\rm MRL}$}  & \colhead{$\sigma_{cross-check}$}
}
\startdata
NGC 0104  & 0.011  &  0.011 \\
NGC 5024  & 0.024  &  0.013 \\
NGC 5272  & 0.011  &  0.012 \\
NGC 7089  & 0.018  &  0.006

\enddata
\end{deluxetable}

\clearpage

\begin{deluxetable}{cr|c}
\tabletypesize{\scriptsize}
%\rotate
\tablecaption{Summary of $M_{\rm F606W}^{\rm TO}$ uncertainty contributions \label{stddevs}}
\tablewidth{0pt}
\tablehead{\colhead{Contribution} & \colhead{mag.} & \colhead{} 
}
\startdata
$\sigma_{\rm MRL}$     &  $\sim$0.01-0.04 &  \\
$\sigma_{\rm MSF}$     &  $\sim$0.01-0.06 & \\
\hline
$\sigma_{\rm NUM}$    &  0.016  &  \\
$\sigma_{\rm RED}$     & 0.020   & Quadratic sum \\
$\sigma_{\rm BIN}$       & 0.0059 &  0.026 mag. \\
$\sigma_{\rm CROW}$ & 0.0015  &
\enddata
\end{deluxetable}

\clearpage

\begin{deluxetable}{lccccccccc}
\tabletypesize{\scriptsize}
\rotate
\tablecaption{Globular cluster relative ages \label{GCages}}
\tablewidth{0pt}
\tablehead{
\colhead{Name} & \colhead{[M/H]$_{\rm ZW}$} & \colhead{[M/H]$_{\rm CG}$} & 
\colhead{$M_{\rm F606W}^{\rm TO}$} & \colhead{D07$_{\rm ZW}$} & \colhead{D07$_{\rm CG}$}  & 
\colhead{P04$_{\rm CG}$}  & \colhead{B94$_{\rm CG}$}  & \colhead{G00$_{\rm CG}$} & Group }
\startdata
NGC 0104 &   -0.57  &  -0.64   & 4.13 $\pm$ 0.07  &  0.96 $\pm$ 0.07	&   1.02 $\pm$ 0.07  &  1.07 $\pm$ 0.08  & 1.02  $\pm$ 0.07  & 1.05  $\pm$ 0.09 & old \\
NGC 0288 &   -1.18  &  -0.92   & 3.82 $\pm$ 0.04  &  0.88 $\pm$ 0.04	&   0.83 $\pm$ 0.03  &  0.83 $\pm$ 0.03  & 0.82  $\pm$ 0.03  & 0.83  $\pm$ 0.03 & young \\
NGC 0362 &   -1.11  &  -0.87   & 3.81 $\pm$ 0.05  &  0.84 $\pm$ 0.04	&   0.81 $\pm$ 0.04  &  0.80 $\pm$ 0.04  & 0.81  $\pm$ 0.04  & 0.81  $\pm$ 0.05 & young \\
NGC 1261 &   -1.10  &  -0.86   & 3.80 $\pm$ 0.05  &  0.83 $\pm$ 0.04	&   0.80 $\pm$ 0.04  &  0.79 $\pm$ 0.04  & 0.80  $\pm$ 0.04  & 0.79  $\pm$ 0.05 & young \\
NGC 1851 &   -1.01  &  -0.81   & 3.79 $\pm$ 0.05  &  0.80 $\pm$ 0.04	&   0.78 $\pm$ 0.04  &  0.77 $\pm$ 0.04  & 0.78  $\pm$ 0.04  & 0.75  $\pm$ 0.04 & young \\
NGC 2298 &   -1.69  &  -1.49   & 3.80 $\pm$ 0.05  &  1.01 $\pm$ 0.05	&   0.99 $\pm$ 0.05  &  0.99 $\pm$ 0.05  & 0.99  $\pm$ 0.05  & 0.99  $\pm$ 0.05 & old \\
NGC 2808 &   -1.14  &  -0.89   & 3.86 $\pm$ 0.03  &  0.91 $\pm$ 0.03	&   0.85 $\pm$ 0.03  &  0.85 $\pm$ 0.03  & 0.85  $\pm$ 0.03  & 0.85  $\pm$ 0.02 & young \\
NGC 3201 &   -1.31  &  -1.02   & 3.75 $\pm$ 0.04  &  0.86 $\pm$ 0.03	&   0.80 $\pm$ 0.03  &  0.79 $\pm$ 0.03  & 0.80  $\pm$ 0.02  & 0.81  $\pm$ 0.03 & young \\
NGC 4147 &   -1.55  &  -1.28   & 3.74 $\pm$ 0.04  &  0.93 $\pm$ 0.04	&   0.89 $\pm$ 0.04  &  0.88 $\pm$ 0.04  & 0.89  $\pm$ 0.04  & 0.89  $\pm$ 0.03 & young \\
NGC 4590 &   -1.89  &  -1.78   & 3.61 $\pm$ 0.05  &  0.89 $\pm$ 0.03	&   0.90 $\pm$ 0.04  &  0.89 $\pm$ 0.04  & 0.90  $\pm$ 0.04  & 0.91  $\pm$ 0.04 & old \\
NGC 4833 &   -1.70  &  -1.49   & 3.80 $\pm$ 0.05  &  1.01 $\pm$ 0.05	&   0.98 $\pm$ 0.05  &  0.98 $\pm$ 0.04  & 0.98  $\pm$ 0.05  & 0.98  $\pm$ 0.05 & old \\
NGC 5024 &   -1.82  &  -1.64   & 3.76 $\pm$ 0.05  &  1.01 $\pm$ 0.05	&   0.99 $\pm$ 0.05  &  0.99 $\pm$ 0.04  & 1.00  $\pm$ 0.05  & 0.99  $\pm$ 0.05 & old \\
NGC 5053 &   -1.88  &  -1.76   & 3.69 $\pm$ 0.05  &  0.95 $\pm$ 0.04	&   0.96 $\pm$ 0.04  &  0.97 $\pm$ 0.04  & 0.96  $\pm$ 0.05  & 0.96  $\pm$ 0.04 & old \\
NGC 5139 &   -1.37  &  -1.13   & 3.82 $\pm$ 0.05  &  0.94 $\pm$ 0.05	&   0.90 $\pm$ 0.05  &  0.89 $\pm$ 0.05  & 0.90  $\pm$ 0.05  & 0.89  $\pm$ 0.05 & young \\
NGC 5272 &   -1.44  &  -1.12   & 3.83 $\pm$ 0.04  &  0.96 $\pm$ 0.04	&   0.89 $\pm$ 0.04  &  0.88 $\pm$ 0.04  & 0.90  $\pm$ 0.03  & 0.89  $\pm$ 0.04 & young \\
NGC 5286 &   -1.48  &  -1.19   & 3.89 $\pm$ 0.05  &  1.04 $\pm$ 0.04	&   0.98 $\pm$ 0.04  &  0.98 $\pm$ 0.05  & 0.97  $\pm$ 0.05  & 0.98  $\pm$ 0.05 & old \\
NGC 5466 &   -2.00  &  -1.98   & 3.74 $\pm$ 0.05  &  1.03 $\pm$ 0.05	&   1.06 $\pm$ 0.05  &  1.06 $\pm$ 0.06  & 1.08  $\pm$ 0.03  & 1.07  $\pm$ 0.05 & old \\
NGC 5904 &   -1.16  &  -0.90   & 3.83 $\pm$ 0.03  &  0.88 $\pm$ 0.03	&   0.83 $\pm$ 0.03  &  0.83 $\pm$ 0.03  & 0.83  $\pm$ 0.03  & 0.83  $\pm$ 0.02 & young \\
NGC 5927 &   -0.18  &  -0.50   & 4.15 $\pm$ 0.07  &  0.88 $\pm$ 0.06	&   0.99 $\pm$ 0.07  &  1.03 $\pm$ 0.08  & 0.98  $\pm$ 0.07  & 1.01  $\pm$ 0.11 & old \\
NGC 5986 &   -1.43  &  -1.13   & 3.88 $\pm$ 0.05  &  1.01 $\pm$ 0.04	&   0.95 $\pm$ 0.04  &  0.94 $\pm$ 0.05  & 0.93  $\pm$ 0.04  & 0.95  $\pm$ 0.05 & old \\
NGC 6093 &   -1.53  &  -1.25   & 3.87 $\pm$ 0.05  &  1.03 $\pm$ 0.04	&   0.98 $\pm$ 0.04  &  0.98 $\pm$ 0.05  & 0.97  $\pm$ 0.05  & 0.98  $\pm$ 0.05 & old \\
NGC 6101 &   -1.73  &  -1.54   & 3.78 $\pm$ 0.04  &  1.00 $\pm$ 0.04	&   0.98 $\pm$ 0.04  &  0.98 $\pm$ 0.04  & 0.98  $\pm$ 0.04  & 0.98  $\pm$ 0.04 & old \\
NGC 6121 &   -1.05  &  -0.83   & 4.01 $\pm$ 0.05  &  1.02 $\pm$ 0.05	&   0.98 $\pm$ 0.05  &  0.99 $\pm$ 0.07  & 0.97  $\pm$ 0.05  & 0.98  $\pm$ 0.05 & old \\
NGC 6144 &   -1.71  &  -1.52   & 3.88 $\pm$ 0.05  &  1.10 $\pm$ 0.05	&   1.08 $\pm$ 0.05  &  1.07 $\pm$ 0.06  & 1.05  $\pm$ 0.05  & 1.07  $\pm$ 0.05 & old \\
NGC 6171 &   -0.87  &  -0.81   & 4.14 $\pm$ 0.05  &  1.08 $\pm$ 0.05	&   1.09 $\pm$ 0.06  &  1.11 $\pm$ 0.06  & 1.07  $\pm$ 0.06  & 1.13  $\pm$ 0.08 & old \\
NGC 6205 &   -1.41  &  -1.11   & 3.84 $\pm$ 0.04  &  0.97 $\pm$ 0.04	&   0.91 $\pm$ 0.04  &  0.90 $\pm$ 0.04  & 0.90  $\pm$ 0.02  & 0.90  $\pm$ 0.04 & young \\
NGC 6218 &   -1.18  &  -0.92   & 3.99 $\pm$ 0.04  &  1.05 $\pm$ 0.04	&   0.99 $\pm$ 0.03  &  1.00 $\pm$ 0.05  & 0.99  $\pm$ 0.04  & 0.99  $\pm$ 0.03 & old \\
NGC 6254 &   -1.33  &  -1.03   & 3.86 $\pm$ 0.04  &  0.96 $\pm$ 0.04	&   0.89 $\pm$ 0.04  &  0.88 $\pm$ 0.04  & 0.90  $\pm$ 0.03  & 0.88  $\pm$ 0.04 & young \\
NGC 6304 &   -0.24  &  -0.52   & 4.19 $\pm$ 0.07  &  0.93 $\pm$ 0.06	&   1.06 $\pm$ 0.08  &  1.09 $\pm$ 0.09  & 1.05  $\pm$ 0.07  & 1.09  $\pm$ 0.10 & old \\
NGC 6341 &   -2.02  &  -1.94   & 3.72 $\pm$ 0.04  &  1.02 $\pm$ 0.04	&   1.03 $\pm$ 0.04  &  1.03 $\pm$ 0.05  & 1.05  $\pm$ 0.04  & 1.03  $\pm$ 0.05 & old \\
NGC 6352 &   -0.36  &  -0.56   & 4.13 $\pm$ 0.07  &  0.90 $\pm$ 0.06	&   0.99 $\pm$ 0.07  &  1.03 $\pm$ 0.08  & 0.99  $\pm$ 0.07  & 1.02  $\pm$ 0.10 & old \\
NGC 6362 &   -0.96  &  -0.85   & 4.09 $\pm$ 0.05  &  1.06 $\pm$ 0.05	&   1.06 $\pm$ 0.05  &  1.08 $\pm$ 0.05  & 1.04  $\pm$ 0.03  & 1.07  $\pm$ 0.06 & old \\
NGC 6366 &   -0.44  &  -0.59   & 4.16 $\pm$ 0.12  &  0.95 $\pm$ 0.07	&   1.04 $\pm$ 0.13  &  1.08 $\pm$ 0.16  & 1.04  $\pm$ 0.13  & 1.07  $\pm$ 0.18 & old \\
NGC 6388 &   -0.60  &  -0.63   & 4.05 $\pm$ 0.08  &  0.90 $\pm$ 0.07	&   0.94 $\pm$ 0.08  &  0.95 $\pm$ 0.10  & 0.92  $\pm$ 0.07  & 0.94  $\pm$ 0.10 & old \\
NGC 6397 &   -1.72  &  -1.54   & 3.79 $\pm$ 0.04  &  1.01 $\pm$ 0.04	&   0.99 $\pm$ 0.04  &  0.99 $\pm$ 0.04  & 0.99  $\pm$ 0.04  & 0.99  $\pm$ 0.04 & old \\
NGC 6441 &   -0.45  &  -0.46   & 4.04 $\pm$ 0.08  &  0.85 $\pm$ 0.06	&   0.88 $\pm$ 0.07  &  0.90 $\pm$ 0.07  & 0.88  $\pm$ 0.05  & 0.85  $\pm$ 0.09 & old \\
NGC 6496 &   -0.36  &  -0.56   & 4.11 $\pm$ 0.07  &  0.88 $\pm$ 0.06	&   0.97 $\pm$ 0.07  &  1.00 $\pm$ 0.09  & 0.95  $\pm$ 0.07  & 0.98  $\pm$ 0.10 & old \\
NGC 6535 &   -1.56  &  -1.29   & 3.68 $\pm$ 0.11  &  0.87 $\pm$ 0.04	&   0.82 $\pm$ 0.09  &  0.82 $\pm$ 0.09  & 0.81  $\pm$ 0.09  & 0.83  $\pm$ 0.08 & young \\
NGC 6541 &   -1.57  &  -1.31   & 3.88 $\pm$ 0.04  &  1.06 $\pm$ 0.04	&   1.01 $\pm$ 0.04  &  1.02 $\pm$ 0.04  & 1.00  $\pm$ 0.03  & 1.01  $\pm$ 0.04 & old \\
NGC 6584 &   -1.32  &  -1.10   & 3.82 $\pm$ 0.03  &  0.92 $\pm$ 0.03	&   0.88 $\pm$ 0.03  &  0.87 $\pm$ 0.03  & 0.89  $\pm$ 0.03  & 0.87  $\pm$ 0.03 & young \\
NGC 6624 &   -0.36  &  -0.56   & 4.12 $\pm$ 0.07  &  0.89 $\pm$ 0.06	&   0.98 $\pm$ 0.07  &  1.01 $\pm$ 0.08  & 0.98  $\pm$ 0.07  & 1.00  $\pm$ 0.10 & old \\
NGC 6637 &   -0.58  &  -0.64   & 4.13 $\pm$ 0.07  &  0.96 $\pm$ 0.07	&   1.02 $\pm$ 0.07  &  1.07 $\pm$ 0.08  & 1.02  $\pm$ 0.07  & 1.05  $\pm$ 0.09 & old \\
NGC 6652 &   -0.75  &  -0.76   & 4.07 $\pm$ 0.06  &  0.97 $\pm$ 0.06	&   1.01 $\pm$ 0.06  &  1.03 $\pm$ 0.08  & 1.01  $\pm$ 0.06  & 1.03  $\pm$ 0.08 & old \\
NGC 6656 &   -1.53  &  -1.27   & 3.87 $\pm$ 0.05  &  1.03 $\pm$ 0.05	&   0.99 $\pm$ 0.05  &  0.99 $\pm$ 0.05  & 0.98  $\pm$ 0.05  & 0.98  $\pm$ 0.05 & old \\
NGC 6681 &   -1.42  &  -1.13   & 3.93 $\pm$ 0.04  &  1.07 $\pm$ 0.04	&   1.00 $\pm$ 0.04  &  1.01 $\pm$ 0.04  & 1.00  $\pm$ 0.03  & 1.01  $\pm$ 0.04 & old \\
NGC 6715 &   -1.32  &  -1.03   & 3.79 $\pm$ 0.03  &  0.90 $\pm$ 0.03	&   0.84 $\pm$ 0.03  &  0.83 $\pm$ 0.03  & 0.83  $\pm$ 0.04  & 0.84  $\pm$ 0.02 & young \\
NGC 6717 &   -1.11  &  -0.87   & 4.06 $\pm$ 0.05  &  1.09 $\pm$ 0.05	&   1.03 $\pm$ 0.05  &  1.07 $\pm$ 0.05  & 1.03  $\pm$ 0.04  & 1.04  $\pm$ 0.06 & old \\
NGC 6723 &   -0.90  &  -0.82   & 4.07 $\pm$ 0.05  &  1.02 $\pm$ 0.05	&   1.02 $\pm$ 0.05  &  1.06 $\pm$ 0.05  & 1.02  $\pm$ 0.04  & 1.04  $\pm$ 0.06 & old \\
NGC 6752 &   -1.32  &  -1.02   & 3.88 $\pm$ 0.03  &  0.98 $\pm$ 0.03	&   0.92 $\pm$ 0.04  &  0.91 $\pm$ 0.03  & 0.91  $\pm$ 0.02  & 0.92  $\pm$ 0.04 & young \\
NGC 6779 &   -1.72  &  -1.50   & 3.89 $\pm$ 0.05  &  1.10 $\pm$ 0.05	&   1.07 $\pm$ 0.05  &  1.06 $\pm$ 0.06  & 1.05  $\pm$ 0.04  & 1.06  $\pm$ 0.04 & old \\
NGC 6809 &   -1.58  &  -1.32   & 3.83 $\pm$ 0.04  &  1.01 $\pm$ 0.04	&   0.96 $\pm$ 0.04  &  0.96 $\pm$ 0.05  & 0.95  $\pm$ 0.04  & 0.96  $\pm$ 0.05 & old \\
NGC 6838 &   -0.44  &  -0.59   & 4.19 $\pm$ 0.07  &  0.98 $\pm$ 0.08	&   1.07 $\pm$ 0.08  &  1.10 $\pm$ 0.10  & 1.06  $\pm$ 0.08  & 1.11  $\pm$ 0.10 & old \\
NGC 6934 &   -1.32  &  -1.10   & 3.81 $\pm$ 0.04  &  0.91 $\pm$ 0.04	&   0.87 $\pm$ 0.04  &  0.86 $\pm$ 0.04  & 0.88  $\pm$ 0.04  & 0.86  $\pm$ 0.03 & young \\
NGC 6981 &   -1.28  &  -0.99   & 3.82 $\pm$ 0.03  &  0.91 $\pm$ 0.02	&   0.85 $\pm$ 0.02  &  0.84 $\pm$ 0.02  & 0.85  $\pm$ 0.03  & 0.85  $\pm$ 0.02 & young \\
NGC 7078 &   -1.91  &  -1.80   & 3.74 $\pm$ 0.05  &  1.00 $\pm$ 0.04	&   1.01 $\pm$ 0.04  &  1.01 $\pm$ 0.04  & 1.02  $\pm$ 0.04  & 1.01  $\pm$ 0.04 & old \\
NGC 7089 &   -1.39  &  -1.09   & 3.86 $\pm$ 0.04  &  0.98 $\pm$ 0.04	&   0.92 $\pm$ 0.04  &  0.91 $\pm$ 0.04  & 0.91  $\pm$ 0.03  & 0.92  $\pm$ 0.05 & young \\
NGC 7099 &   -1.83  &  -1.70   & 3.76 $\pm$ 0.05  &  1.01 $\pm$ 0.04	&   1.01 $\pm$ 0.05  &  1.01 $\pm$ 0.04  & 1.01  $\pm$ 0.04  & 1.01  $\pm$ 0.04 & old \\
ARP 2    &   -1.52  &  -1.23   & 3.74 $\pm$ 0.07  &  0.91 $\pm$ 0.06	&   0.85 $\pm$ 0.06  &  0.84 $\pm$ 0.06  & 0.84  $\pm$ 0.06  & 0.85  $\pm$ 0.05 & young \\
E3       &   -0.66  &  -0.69   & 4.09 $\pm$ 0.11  &  0.95 $\pm$ 0.05	&   1.00 $\pm$ 0.11  &  1.02 $\pm$ 0.13  & 1.00  $\pm$ 0.10  & 1.02  $\pm$ 0.15 & old \\
LYNG\AA7 &   -0.48  &  -0.50   & 4.27 $\pm$ 0.12  &  1.09 $\pm$ 0.08	&   1.13 $\pm$ 0.14  &  1.18 $\pm$ 0.14  & 1.10  $\pm$ 0.15  & 1.19  $\pm$ 0.18 & old \\
PAL 1    &   -0.46  &  -0.56   & 3.59 $\pm$ 0.14  &  0.54 $\pm$ 0.06	&   0.57 $\pm$ 0.09  &  0.59 $\pm$ 0.05  & 0.63  $\pm$ 0.07  & 0.55  $\pm$ 0.10 & young \\
PAL 12   &   -0.68  &  -0.69   & 3.73 $\pm$ 0.11  &  0.67 $\pm$ 0.04	&   0.69 $\pm$ 0.09  &  0.69 $\pm$ 0.08  & 0.72  $\pm$ 0.07  & 0.68  $\pm$ 0.07 & young \\
TERZAN7  &   -0.44  &  -0.42   & 3.63 $\pm$ 0.07  &  0.55 $\pm$ 0.04	&   0.57 $\pm$ 0.04  &  0.57 $\pm$ 0.05  & 0.63  $\pm$ 0.04  & 0.51  $\pm$ 0.05 & young \\
TERZAN8  &   -1.75  &  -1.58   & 3.73 $\pm$ 0.05  &  0.96 $\pm$ 0.04	&   0.95 $\pm$ 0.04  &  0.95 $\pm$ 0.04  & 0.94  $\pm$ 0.05  & 0.95  $\pm$ 0.04 & old 
\enddata
\end{deluxetable}

\end{document}